\documentclass[conference]{IEEEtran}
\IEEEoverridecommandlockouts
\usepackage{cite}
\usepackage{amsmath,amssymb,amsfonts}
\usepackage{algorithmic}
\usepackage{textcomp}
\usepackage{xcolor}
\def\BibTeX{{\rm B\kern-.05em{\sc i\kern-.025em b}\kern-.08em
    T\kern-.1667em\lower.7ex\hbox{E}\kern-.125emX}}
    
\usepackage{floatrow}
\newfloatcommand{capbtabbox}{table}[][\FBwidth] 
    
    \usepackage{diagbox}
\usepackage{tikz}
\usepackage{amsmath}
\usepackage{amsthm}

\usepackage{soul,color}
\usepackage{etoolbox}
\usepackage{wrapfig}
\usepackage{wrapfig,lipsum,booktabs}
\usepackage{amsfonts}

\usepackage{mathtools}
\usepackage{graphicx}
\usepackage{makecell}
\usepackage[ruled,linesnumbered,vlined]{algorithm2e}
\usepackage{siunitx,  % for 'S' column type
            booktabs, % for well-spaced horiz. lines
            caption} 
            
\usepackage{amssymb,graphicx}

\usepackage{adjustbox}

\usepackage{hyperref}
\hypersetup{
    colorlinks=true,
    linkcolor=blue,
    filecolor=magenta,      
    urlcolor=cyan,
    pdftitle={Overleaf Example},
    pdfpagemode=FullScreen,
    }

\usepackage{multirow}
\usepackage{color, colortbl}
\definecolor{Gray}{gray}{0.9}
\definecolor{Red}{rgb}{1,0.7,0.7 }
\definecolor{Blue}{rgb}{0.6,0.9,0.97 }
\definecolor{Green}{rgb}{0.55,0.92,0.55 }
\usepackage{pifont}%
\newcommand{\cmark}{\ding{51}}%
\newcommand{\xmark}{\ding{55}}%

\makeatletter
\newcommand\bigDiamond{\mathop{\mathpalette\bigDi@mond\relax}}
\newcommand\bigDi@mond[2]{%
  \vcenter{\hbox{\m@th
    \scalebox{\ifx#1\displaystyle 2\else1.2\fi}{$#1\Diamond$}%
  }}%
}
\newcommand\bigLozenge{\mathop{\mathpalette\bigL@zenge\relax}}
\newcommand\bigL@zenge[2]{%
  \vcenter{\hbox{\m@th
    \scalebox{\ifx#1\displaystyle 2\else1.2\fi}{$#1\blacklozenge$}%
  }}%
}
\makeatother

\DeclareMathOperator{\R}{\mathbf{R}}

\DeclarePairedDelimiter{\floor}{\lfloor}{\rfloor}

\begin{document}

% \title{Towards Privacy-preserving Fingerprinting of Relational Databases
% % {\footnotesize \textsuperscript{*}Note: Sub-titles are not captured in Xplore and
% % should not be used}
% % \thanks{Identify applicable funding agency here. If none, delete this.}
% }

\title{Robust Fingerprinting of Genomic Databases} 
% \author{Anonymous Author(s)}

\author{\IEEEauthorblockN{Tianxi Ji}
\IEEEauthorblockA{\textit{Dept.  of ECSE} \\
\textit{Case Western Reserve Univ.}\\
%City, Country \\
txj116@case.edu}
\and
\IEEEauthorblockN{Erman Ayday$^\dagger$}\thanks{$^\dagger$Corresponding author.}
\IEEEauthorblockA{\textit{Dept.   of CSDS} \\
\textit{Case Western Reserve Univ.}\\
% City, Country \\
exa208@case.edu}
\and
\IEEEauthorblockN{Emre Yilmaz$^*$}\thanks{$^*$Part of this research was undertaken while the author was at Case Western Reserve University.}
\IEEEauthorblockA{\textit{Dept.  of CSET} \\
\textit{Univ. of Houston-Downtown}\\
% City, Country \\
yilmaze@uhd.edu}
\and
\IEEEauthorblockN{Pan Li}
\IEEEauthorblockA{\textit{Dept.  of ECSE} \\
\textit{Case Western Reserve Univ.}\\
% City, Country \\
lipan@case.edu}
% \and
% \IEEEauthorblockN{5\textsuperscript{th} Given Name Surname}
% \IEEEauthorblockA{\textit{dept. name of organization (of Aff.)} \\
% \textit{name of organization (of Aff.)}\\
% City, Country \\
% email address or ORCID}
% \and
% \IEEEauthorblockN{6\textsuperscript{th} Given Name Surname}
% \IEEEauthorblockA{\textit{dept. name of organization (of Aff.)} \\
% \textit{name of organization (of Aff.)}\\
% City, Country \\
% email address or ORCID}
\thanks{Research reported in this publication was partly supported by the National Library Of Medicine of the National Institutes of Health under Award Number R01LM013429 and by the National Science Foundation (NSF) under grant number 2050410. A peer-reviewed version of this paper will appear in the 
30th International Conference on Intelligent Systems for Molecular Biology  (ISMB'22). Copyright will be transferred without notice. There are some minor editorial differences between the
two versions.}
}

\maketitle
\thispagestyle{plain}
\pagestyle{plain}
\begin{abstract}
% \hl{remove the aim of utility?}

Relational database fingerprinting (a   steganography technique that randomly changes database entries)  
has been widely used to discourage unauthorized redistribution of databases by providing means to identify the source of data leakages.  However, there is no fingerprinting scheme aiming at achieving liability guarantees when sharing genomic databases. Thus, we are motivated to fill in this gap by devising a vanilla fingerprinting scheme specifically for genomic databases.  Moreover, since  malicious genomic database recipients may   compromise the embedded fingerprint (distort the steganographic marks, i.e., the embedded fingerprint bit string) by launching effective correlation attacks which leverage the intrinsic correlations among genomic data (e.g., Mendel's law and linkage disequilibrium), we also augment the vanilla scheme by developing mitigation techniques to achieve robust fingerprinting of genomic databases against correlation attacks.

Via experiments using a real-world genomic database, we first show that correlation attacks against fingerprinting schemes for genomic databases are very powerful. In particular,  the correlation attacks can distort more than half of the fingerprint bits by causing a small utility loss (e.g., database accuracy and consistency of SNP-phenotype associations measured via p-values). Next, we experimentally show that the correlation attacks can be effectively mitigated by our proposed mitigation techniques. We validate  that the attacker can hardly compromise a large portion of the fingerprint bits even if it pays a higher cost in terms of degradation of the database utility. 
For example, with around $24\%$ loss in accuracy and $20\%$ loss in the consistency of SNP-phenotype associations, the attacker can only distort about $30\%$ fingerprint bits, which is insufficient for it to avoid being accused.  We also show that the proposed mitigation techniques also preserve the utility of the shared genomic databases, e.g., the mitigation techniques   only lead to around $3\%$ loss in accuracy.\footnote{Availability and implementation: \url{https://github.com/xiutianxi/robust-genomic-fp-github}}

% \textbf{Availability and implementation: \url{https://github.com/xiutianxi/robust-genomic-fp-github}\footnote{Research reported in this publication was partly supported by the National Library Of Medicine of the National Institutes of Health under Award Number R01LM013429 and by the National Science Foundation (NSF) under grant number 2050410.}} 
\end{abstract}

% \begin{IEEEkeywords}
% database, privacy, liability, fingerprinting
% \end{IEEEkeywords}

% \hl{TJ: do we need to include CCS concepts and keywords?}

\section{Introduction}\label{sec:introduction}

Genomic database sharing is critical in modern biomedical research,  clinical practice, and customized  healthcare. However, it is generally not viable due to the copyright and intellectual property    concerns from the   database  owners. In other words, the requirements of   copyright protection and anti-piracy may prevent genomic data holders from sharing their data, which may hinder the progress of  cooperative scientific research.

Digital fingerprinting is a technology that allows to claim 
copyright, deter illegal redistribution, and identify the source of data breaches (i.e., the guilty party who is responsible for the leakage) by embedding a unique mark into each shared copy of a digital object. 
%Unlike digital watermarking, in fingerprinting, the embedded mark must be unique to detect the guilty party who is responsible for the leakage. 
Although the most prominent usage of fingerprinting is for multimedia   \cite{cox1997secure,cox2002digital,johnson2001information}, fingerprinting techniques for databases have also been developed \cite{li2005fingerprinting,guo2006fingerprinting,liu2004block,lafaye2008watermill}. These techniques change database entries at different positions when sharing a database copy with a new service provider (SP).  If the SP shares its received copy without authorization, the database owner can use the inserted fingerprints to  hold the guilty SP responsible.

%\vspace{-4mm}
\subsection{Challenges in Genomic Database Fingerprinting}
%\vspace{-2mm}
Existing fingerprinting schemes for databases have been developed to embed fingerprints in continuous  numerical entries (floating points) in relational databases, e.g.,  \cite{li2005fingerprinting,guo2006fingerprinting,li2003constructing}. Whereas, fingerprinting discrete (or categorical) values is more difficult, as the number of possible values (or instances) for a data point is much fewer. Hence, in such databases, a small change in   data points (as a fingerprint) can significantly affect the utility. Fingerprinting becomes even more challenging when it comes to a genomic databases, which contain even fewer values, e.g., 4 instances (A, G, C, T) when considering nucleobases, and 3 instances (0, 1, 2) when considering number of minor alleles for each single nucleotide polymorphism - SNP (Section \ref{sec:system-threat} gives more details about the type of genomic data  considered in this work).   In real world,   leaked genomic databases often end up being sold or publicly shared on the internet \cite{mcgee_ross}. Once that happens, the genomic database owner wants to  find out the traitors who should  be responsible for the data leakage by  extracting the their fingerprints in the leaked databases. Thus, in this paper, we first propose a vanilla genomic database fingerprinting scheme by (i) taking into account 
of the abundant attributes of genomic sequence, and (ii) extending the state-of-the-art database fingerprinting scheme \cite{li2005fingerprinting}. Our  vanilla scheme is more robust against common attacks targeted on database fingerprinting schemes (e.g.,   random bit flipping attack and   subset attack \cite{li2003constructing,agrawal2003watermarking})  than previously developed fingerprinting schemes for generic relational databases such as \cite{li2005fingerprinting,guo2006fingerprinting,liu2004block}, because we can insert   denser fingerprint for each selected genomic data by introducing a new   parameter,   which controls the percentage of fingerprinted entries for   selected rows (see Section \ref{sec:vanilla_insert}). Whereas, \cite{li2005fingerprinting} only fingerprints one attribute for all selected rows. Compared with \cite{li2005fingerprinting}, we also assign higher confidence score  during fingerprint extraction considering that genomic databases usually contains more attributes than generic databases (see Section \ref{sec:vanilla_extract}).

In addition, existing fingerprinting schemes for databases do not consider various inherent correlations between the data records in a database.  In our  previous work \cite{ji_fp_raid2021},   we have shown that a malicious party having a fingerprinted copy of a database can detect and distort the embedded fingerprints using its knowledge about the correlations in the data entries.  Genomic databases contain  even richer row- and column-wise correlations   due to the biological characteristics. In particular, the row-wise correlations arise from (i) the Mendel's law, and (ii) similarities of 
%there exists 3 types of remarkable correlations in genome: (i) hereditary units from parents governed by the Mendel's law, (ii) correlation 
genomes among family members. The column-wise correlations are the  pairwise correlation between genomic data points at different locations (e.g., linkage disequilibrium \cite{naveed2015privacy}).  
In this paper, we use %$\mathrm{Atk_{Mendel}}$,
$\mathrm{Atk_{row}(\mathcal{S})}$ and   $\mathrm{Atk_{col}}(\mathcal{J})$  to represent the   correlation attacks using the row- and column-wise correlations, respectively, where $\mathcal{S}$  and $\mathcal{J}$  denote     the corresponding correlation model and they are assumed to be publicly known (in Section \ref{sec:threat_models}, we     describe  these two attacks in detail). 
%$\mathcal{S}$    denotes  the the malicious SP's prior knowledge    on the correlations between genomic data of family members, and $\mathcal{J}$ denotes its knowledge on the pairwise correlation between SNPs. 
In  Section \ref{sec:geo_experiment}, we consider a real world genomic database and show that by launching %$\mathrm{Atk_{Mendel}}$,
$\mathrm{Atk_{row}(\mathcal{S})}$ and   $\mathrm{Atk_{col}}(\mathcal{J})$ in sequence,  a malicious SP can easily compromise more than half of the bits in a fingerprint string at the cost of only changing about 5\% of the entries in the genomic databases. %, even though the   vanilla scheme developed specially for the genomic databases is already robuster than the one in \cite{li2005fingerprinting}.
As a result, we also need to make the proposed vanilla genomic database fingerprinting scheme be robust against the correlation attacks in order to  lay a solid foundation for genomic data sharing.

% \hl{For example, it is shown that single nucleotide polymorphisms (SNPs)
% may have pairwise correlations between each other (e.g., linkage
% disequilibrium [31]) and an attacker can use such correlations to
% infer the original values of the shared SNPs in the database.}
% %For example, the zip codes are strongly correlated with street names in a demographic database. 
% Similarly, genomes of family members are correlated in a genomic database making common fingerprinting schemes vulnerable to attacks utilizing such correlations. Thus,   to provide robustness against correlation attacks (which utilizes the correlations between attributes and data records to infer the potentially fingerprinted entries), we need to consider  such correlations when developing  fingerprinting schemes for genomic databases. % need to consider such correlations. % in the database.

% \hl{in this paper, is it ok to not distinguish S J and S' J'? jsut refer S and J as publicly know.}

%\vspace{-6mm}
\subsection{Our Solution}
%\vspace{-1.5mm}
In this work,  to address the unique challenges of robust fingerprinting of genomic databases, i.e., mitigating   %$\mathrm{Atk_{Mendel}}$,
$\mathrm{Atk_{row}(\mathcal{S})}$ and   $\mathrm{Atk_{col}}(\mathcal{J})$, we develop   mitigation techniques for each of them, i.e., %$\mathrm{Mtg_{Mendel}}$,
$\mathrm{Mtg_{row}(\mathcal{S})}$ and   $\mathrm{Mtg_{col}}(\mathcal{J})$. %, where $\mathcal{S}$  and $\mathcal{J}$  denote the   database owner's knowledge on the 2nd and 3rd correlations. 
These techniques utilize the correlations among genomic data, i.e., Mendel's law, $\mathcal{S}$, and  $\mathcal{J}$, and they work as post-processing steps for our developed vanilla scheme. Besides, they only modify   non-fingerprinted entries in the genomic databases. Thus, they do not reduce the   robustness of the vanilla scheme. Note that the proposed robust genomic database fingerprinting scheme in this paper is not just a simple application of our previous work \cite{ji_fp_raid2021} for genomic databases, because the correlation models considered in this paper are different compared to the generic models we have \cite{ji_fp_raid2021}, and thus they require new mitigation techniques   to make the fingerprinted genomic databases match the Mendel's law and genome-specific correlations. In Table \ref{table:differences}, we summarize the differences between the proposed robust genomic
database fingerprinting scheme and  previous schemes.

% which we develop in this paper. %\footnote{{\color{red}The proposed robust genomic database fingerprinting scheme in this paper is not just a simple application of our previous scheme \cite{ji_fp_raid2021} on genomic database, because the correlation models considered in this paper are different with that in \cite{ji_fp_raid2021}, and thus  requires  different mitigation techniques.}} 
%Particularly, 

\begin{table}[htb]
% \begin{wraptable}{r}{.55\linewidth}
% %\vspace{-6mm}
    \begin{minipage}{0.9\linewidth}
\begin{adjustbox}{width=1\textwidth}
\begin{tabular}{|l|c|c|c|c|}\hline
% \diagbox{properties}{schemes}
Properties &  \cite{li2005fingerprinting} & \cite{ji_fp_raid2021}  & \thead{this \\paper}\\ 
  \hline
  Flexible density in marked attributes& \xmark&  \xmark&         \cmark      \\ 
  \hline
Higher confidence in extraction  & \xmark&    \xmark&       \cmark   \\
  \hline
Genome-specific correlation & \xmark&    \xmark&       \cmark \\
%   \hline
% perfectly defend   $\mathrm{Atk_{row}(\mathcal{S})}$ & \xmark&    \xmark&      \cmark \\
%   \hline
%   0.09 & 33.59\% &    33.59\% &    34.38\% &    34.38\% &    34.38\% &  35.16\%\\
%   \hline
%   0.1 & 27.34\% &    27.34\% &    27.34\% &    27.34\% &    28.12\% &    28.91\%  \\
    \hline
  \end{tabular}
  \end{adjustbox}
  \end{minipage}
   %\vspace{-5mm}
    \caption{Differences between the robust genomic database fingerprinting and the previous   schemes. \cmark indicates the scheme has a certain property, and \xmark indicates the opposite.}\label{table:differences}
    %\vspace{-8mm}
% \end{wraptable}
\end{table}

 In particular,    $\mathrm{Mtg_{row}(\mathcal{S})}$ is composed of two phases. %(elaborated in Section \ref{sec:mtg_mendel}) 
    First, it checks all fingerprinted genomic data-tuples of family members. If a tuple violates   Mendel's law, the database owner  changes the   non-fingerprinted entries in this tuple to make it compliant with the Mendel's law. Second, it  checks all family sets in the genomic database, calculates the empirical correlations among family members after the vanilla fingerprinting, and changes the non-fingerprinted entries in each family set to push the empirical correlations close to the publicly known model ($\mathcal{S}$) by solving a distance minimization problem. Note that the second phase of $\mathrm{Mtg_{row}(\mathcal{S})}$ is different with the row-wise mitigation developed in our previous work \cite{ji_fp_raid2021}, because the second phase is able to perfectly defend against $\mathrm{Atk_{row}(\mathcal{S})}$ if the objective function of the distance minimization problem reaches to 0. Whereas, our previous  row-wise mitigation technique \cite{ji_fp_raid2021} (formulated as a     set function maximization problem) can only mislead the malicious SP when launching the row-wise correlation attack. More details are deferred to Section \ref{sec:row-defense}.
    
    % \item 
    $\mathrm{Mtg_{col}(\mathcal{J})}$ considers all attributes (columns) of the genomic databases, obtains the empirical marginal distributions after the vanilla fingerprinting, and changes non-fingerprinted entries in each attribute to make the empirical marginal distributions  resemble the marginal distributions obtained by marginalizing the joint distributions provided in $\mathcal{J}$. Similar to our previous work \cite{ji_fp_raid2021}, the database owner selects the  non-fingerprinted entries and modifies their value by solving a linear programming problem %(i.e., entropy regularized Sinkhorn distance minimization ~\cite{courty2016optimal}), 
    which is discussed in Section \ref{sec:col_defense}.
    
    % In particular, to mitigate the column-wise correlation attack, the database owner modifies some of the non-fingerprinted data entries to make the post-processed fingerprinted database have column-wise correlations close to that of her prior knowledge. The data entry modification plans are determined from the solutions to a set of ``optimal transportation'' problems~\cite{courty2016optimal}, {\color{black}each of which transports the mass of the marginal distribution of a specific attribute (column) to make it resemble the reference marginal distribution computed from database owner's prior knowledge while minimizing the transportation cost.
    
% \end{itemize}
In Section \ref{sec:geo_experiment}, we show that by applying %$\mathrm{Mtg_{Mendel}}$,
$\mathrm{Mtg_{row}(\mathcal{S})}$ and   $\mathrm{Mtg_{col}}(\mathcal{J})$ after our proposed vanilla fingerprinting scheme, the malicious SP can 
hardly distort large portion of the fingerprint bits anymore, even if it introduces significant utility loss in the database (e.g., by decreasing database accuracy and making the SNP-phenotype associations less consistent compared to the ground-truth measured on the original - non-fingerprinted - database). % only compromise less than 6\% bits of the fingerprint string even if launches %$\mathrm{Atk_{Mendel}}$,
% $\mathrm{Atk_{row}(\mathcal{S})}$ and   $\mathrm{Atk_{col}}(\mathcal{J})$. 
For example, if the malicious SP compromises   around 24\% accuracy and 20\% consistency in SNP-phenotype associations, it can only distort about 30\% fingerprint bits. This implies that the malicious SP will be   held responsible for the genomic database leakage with high probability \cite{ji_fp_raid2021}.

\textbf{Contributions and Broader Impact.} To the best of our knowledge, our work is the first   to investigate the liability issue when sharing genomic relational databases (i.e., the collection of genomic data of  individuals with the same attributes) and at the same time address the  threats of    correlation attacks due to  the unique biological characteristics.

Our proposed robust genomic database fingerprinting scheme  helps facilitate the development of genomic research which requires   large-scale genomic data analyses, and  is increasingly relying   upon the sharing of   genomic databases with various SPs. The ideas developed in this paper also shed light on sharing other sensitive  biomedical   databases, e.g., electrocardiogram   and electrooculogram data samples, where the data correlations are determined by the spatio-temporal dependency between data records. In this case the correlations can be characterized by  autocorrelation and cross-correlations   or they are modeled as a Markov process.   We will also study these types of databases in the future.

\textbf{Roadmap}.  We review related works on  database fingerprinting schemes in Section \ref{sec:related_work}. In Section \ref{sec:system-threat}, we present the system and threat models, and evaluation metrics. Section \ref{sec:vanilla_fp} introduces the foundation of our scheme, i.e., the vanilla fingerprinting scheme for genomic databases. Then,  in Section \ref{sec:robust_fp}, we consolidate this foundation by developing mitigation techniques against various correlation attacks on   the genomic database. In Section \ref{sec:geo_experiment}, we show the vulnerabilities of the vanilla fingerprinting scheme for genomic databases, and also demonstrate the performance of mitigation techniques from both database utility and fingerprint robustness perspective.  Section \ref{sec:discussion} discusses the limitations of the proposed scheme and points out future research directions.  Finally, Section \ref{sec:conclusion} concludes the paper.

%\vspace{-7mm}
\section{Related Work}
\label{sec:related_work}

The seminal work  of database fingerprinting is proposed by %Agrawal et al. 
\cite{agrawal2003watermarking}, which assumes  that the database consumer can tolerate a small amount of error in the marked databases. Then, based on \cite{agrawal2003watermarking}, some variants have been proposed \cite{li2005fingerprinting,guo2006fingerprinting,liu2004block}. For instance, %Li et al.
~\cite{li2005fingerprinting} develop a database fingerprinting scheme by extending \cite{agrawal2003watermarking} to enable the insertion and extraction of arbitrary bit-strings in relations. However, these works do not consider the correlations among data entries, which makes them vulnerable to correlation attacks ~\cite{ji_fp_raid2021}.  Records in  Genomic databases usually have much stronger correlations caused by  Mendel's law and  linkage disequilibrium, which make the genomic database prone to  correlation attacks.

Recently, some works have explicitly taken the genomic data correlations into account in fingerprinting scheme design. In particular, %Yilmaz et al. 
\cite{yilmaz2020collusion} develop a probabilistic fingerprinting scheme by considering the conditional probabilities between genomic data points of a single individual. %Ayday et al.~
\cite{ayday2019robust} propose an optimization-based fingerprinting scheme for sharing  personal genomic sequential data by jointly considering   collusion attack and  data correlation. However, these two works focus on the genomic data of an individual, instead of a genomic database, where individuals may have kinship relationships. \cite{oksuz2021privacy} develop a watermarking scheme for sequential genomic data based on  belief propagation which 
considers  the privacy of data and the robustness of watermark requirements at the same time.   
\cite{ji_fp_raid2021} propose  mitigation techniques against general correlation attacks targeted on generic  relational     databases and show that the proposed technique can be applied after  any existing relational     database  fingerprinting scheme to achieve robustness against correlation attacks.

However, the above mentioned works cannot be directly applied to genomic database fingerprinting, because   they fail to consider the characteristics that are unique to the genomic data. Particularly, (i) hereditary units governed by the Mendel's law can be utilized by the malicious SP to further infer the potentially fingerprinted locations. (ii) limited values  of genomic data also makes the utility-preserving fingerprinting a challenging task. Thus, in this paper, we first show the vulnerability of genomic database fingerprinting against correlation attacks that take advantage of the Mendel's law and linkage disequilibrium. Then, we discuss how to mitigate the identified attacks in a way that the utility of the database is preserved.

% Our work differs from these works since we focus on  developing robust fingerprint scheme for relational databases, which (i) contain large amount of data records from different individuals, (ii) include both column- and row-wise correlations, and (iii) have different utility requirements.  

%\vspace{-7mm}
\section{System, Threat Model, and Success Metrics}
\label{sec:system-threat}
%\vspace{-1mm}

Now, we discuss the genomic  database fingerprinting system, the considered  various threats, and fingerprint robustness and utility metrics.

%\vspace{-4mm}
\subsection{Genomic Database Fingerprinting System Model}
\label{system_model}
%\vspace{-1mm}

We consider a  database owner (Alice)  with a genomic database (e.g., dbSNP \cite{wheeler2007database}) including  single-nucleotide polymorphisms (SNPs) of a certain population, i.e., each row corresponds to the SNP sequence of a specific individual. Each individual has two alleles for each SNP position, and each of these alleles are inherited from one of their parents.  Thus, each SNP (i.e., each entry of the database) can be represented by the number of its minor alleles as 0, 1, or 2, and can be encoded as ``00'', ``01'', or ``10'', respectively. 
In this paper, we focus on sharing SNP databases, because such databases are critical to many genomic and medical research \cite{mitchell2004discrepancies}, e.g., genome-wide association studies \cite{carlson2003additional}. The techniques developed in this paper can be applied to other types of genomic databases (e.g., ones including nucleotides that may contain 4 values  A, G, C, or T) by simply changing the data coding.

%\vspace{-5mm}
\begin{figure}[htb]
  \begin{center}
     \includegraphics[width=0.9\textwidth]{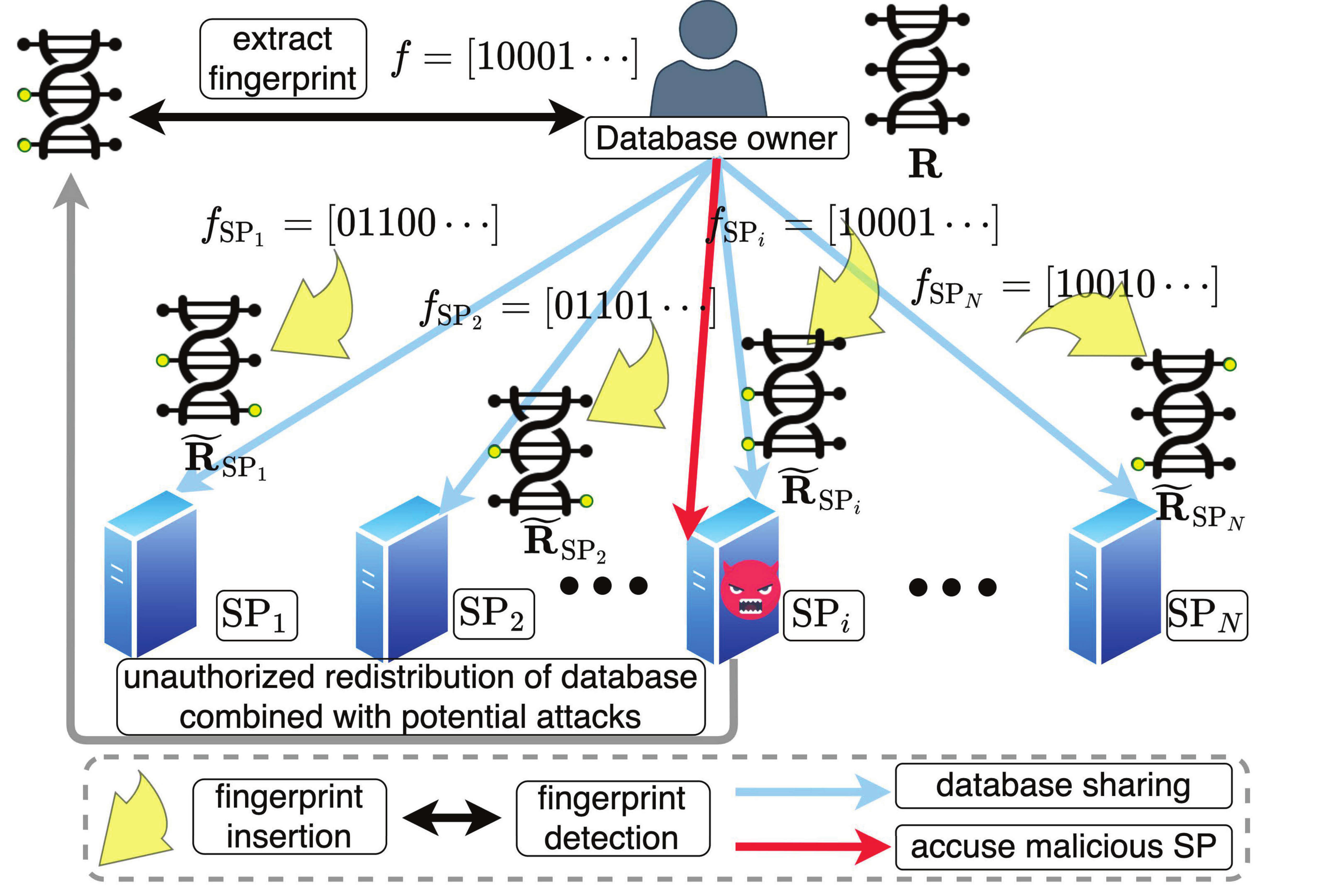}
      \end{center}
      %\vspace{-8mm}
  \caption{\label{fig:fp_general_system} The genomic database fingerprinting system, where Alice adds a unique fingerprint in each copy of her genomic database $\R$ when sharing. The inserted fingerprint will change entries at different locations (indicated by the yellow dots) in $\R$. She is able to identify the malicious SP who pirates and redistributes her database using the customized fingerprint.}
  %\vspace{-6mm}
\end{figure}

% \begin{wrapfigure}{R}{0.53\linewidth}
% \centering
% %\vspace{-10mm}
% \includegraphics[width=1\linewidth]{figures/genomic_fp.eps}
% %\vspace{-9mm}
% \caption{\label{fig:fp_general_system} The genomic database fingerprinting system, where Alice adds a unique fingerprint in each copy of her genomic database $\R$ when sharing. The inserted fingerprint will change entries at different locations (indicated by the yellow dots) in $\R$. She is able to identify the malicious SP who pirates and redistributes her database using the customized fingerprint.}
% %\vspace{-8mm}
% \end{wrapfigure}

We present the  system model for genomic database fingerprinting  in Figure \ref{fig:fp_general_system}. We denote the genomic database owned by Alice as $\mathbf{R}$. When Alice wants to share the database with various 
medical service providers (SPs), she includes a unique fingerprint   in each copy of her database. The fingerprint bit-string customized for each SP is a randomly generated binary bit-string (elaborated in Section \ref{sec:vanilla_insert}). The fingerprint  essentially changes different entries in $\mathbf{R}$ at various positions (indicated by the yellow dots).  
The fingerprint generated  for the $i$th SP  ($\mathrm{SP}_i$) is  $f_{\mathrm{SP}_i}$, and the fingerprinted genomic database received by $\mathrm{SP}_i$ is represented as $\widetilde{\R}_{f_{\mathrm{SP}_i}}$. We also use $\widetilde{\mathbf{R}}$ to represent  a generic fingerprinted genomic database.  %Table \ref{main_notations} lists the frequently used notations in the paper. 

In real world applications, some of the SPs may be malicious (e.g., $\mathrm{SP}_i$ in Figure \ref{fig:fp_general_system}) who will redistribute its received genomic database copy after conducting certain attacks (discussed in Section \ref{sec:threat_models}) on top of it.

\noindent\textbf{System with Vanilla Fingerprinting.}  If $\mathrm{SP}_i$ compromise $\widetilde{\R}_{f_{\mathrm{SP}_i}}$ via random bit flipping attack, Alice 
is able to identify $\mathrm{SP}_i$ as the traitor by extracting a large portion of its fingerprint from the leaked database by only using the   proposed vanilla fingerprint scheme (Section \ref{sec:vanilla_fp}). However, if $\mathrm{SP}_i$  launches correlation-based attacks on $\widetilde{\R}_{f_{\mathrm{SP}_i}}$, it can avoid being accused of data leakage with large probability.

\noindent\textbf{System with Robust Fingerprinting.} The correlation-based attacks can be effectively mitigated, if $\widetilde{\R}_{f_{\mathrm{SP}_i}}$ is generated by adopting our proposed robust fingerprinting  scheme, and thus, $\mathrm{SP}_i$ will still be held responsible for illegal redistribution. We will empirically evaluate these using a real world genomic database in Section \ref{sec:geo_experiment}.

\subsection{Threat Model}
\label{sec:threat_models}
%\vspace{-1mm}
Fingerprinted databases are subject to various attacks summarized in the following. %In Figure \ref{fig:system_and_attacks}(b), we show some representative ones that are studied in this paper. 
Note that in all considered attacks, a malicious SP can change/modify most of the entries in  $\widetilde{\mathbf{R}}$ to distort the fingerprint (and to avoid being accused). However, such a pirated database will have significantly poor utility measured in terms of database accuracy and consistency of SNP-phenotype association (see Section \ref{sec:utility_metrics}). Thus, a rational  SP will try to get away with making pirated copies of the genomic databases by  changing  as few entries as possible  in order to maintain high utility for the pirated database and gain illegal profit.

% \hl{poor utility, e.g., sction...}. % as shown in \cite{ji_fp_raid2021}.  %{\color{black}As discussed in Section \ref{sec:vanilla}, we let the vanilla fingerprint scheme only change the LSBs of data entries to preserve data utility. Thus, all considered attacks also change the LSBs of the selected entries in $\widetilde{\mathbf{R}}$ to distort the fingerprint.}

% Thus, in this work, to preserve the database utility, we  assume that some of the  attacks just change the LSB of the selected entries in $\widetilde{\mathbf{R}}$ to distort the fingerprint. As we show in Appendix \ref{sec:utility_LSB}, changing the LSB of the database entries results in the least utility loss.

\noindent\textbf{Random Bit Flipping Attack.} In this attack, to pirate a database, a malicious SP   selects  random entries in its received copy of the genomic database  %$\widetilde{\mathbf{R}}$ 
and flips their bit values \cite{agrawal2003watermarking}. For example, a SNP value 2 (``10'') becomes 0 (``00'') after the attack.%The flipped entries are still in the domain of the corresponding attributes. 
% The considered vanilla fingerprint scheme is robust against this attack \cite{li2005fingerprinting} as shown in Figure~\ref{fig:system_and_attacks}(b)(i),   Alice shares fingerprinted copies of her database by only applying  $\mathrm{FP}$. If a malicious SP ($\mathrm{SP}_i$) tries to distort the fingerprint in $\widetilde{\mathbf{R}}(\mathrm{FP},\emptyset,\emptyset)$ using the random bit flipping attack (i.e., $\mathrm{Atk_{rnd}}$), and then   redistributes it, Alice can still detect $\mathrm{SP}_i$'s fingerprint in the pirated  copy with a high probability, and correctly accuse $\mathrm{SP}_i$ of data leakage.
% In fact, to distort more than half of the fingerprint bits, $\mathrm{Atk_{rnd}}$ needs to change more than $83\%$ of the database entries (as we will show in Section \ref{sec: int_attack_census}), which compromise the database utilities significantly. 

% \noindent\textbf{Subset and Superset Attacks.}
% In subset attack, a malicious SP generates a pirated copy of $\widetilde{\mathbf{R}}$ by randomly selecting  data records from it.   Superset attack is the dual attack of subset attack. In this attack, the malicious SP mixes $\widetilde{\mathbf{R}}$ with other databases to create a pirated  one. These two attacks are considered to be  weak attacks. For example, for subset attack, to compromise just one specific  bit in the inserted  fingerprint bit-string,  the malicious SP must exclude all records that are marked by that  bit~\cite{li2005fingerprinting}.

\noindent\textbf{Row-wise correlation attack  $\mathrm{Atk_{row}}(\mathcal{S})$.}
As discussed in Section \ref{sec:introduction},  a malicious SP may utilize Mendel's law and similarities among family members' genomes to infer the potentially changed loci in the fingerprinted database. Thus, we assume that the malicious SP has access to the sets of families in the database as well as the genome similarities (denoted as $\mathcal{S}$) among family members in each set. Note that this is a valid assumption, because quite a few works have shown that kinship or familial relationships from SNP genotyping data can be inferred with very high accuracy for small and medium size groups (e.g., dozens or hundreds of individuals \cite{goudet2018estimate,park2013inference}, large size populations \cite{wang2017efficient}, or even worldwide \cite{li2008worldwide}) or such information can be obtained directly from the metadata.

Thus, upon receiving a fingerprinted copy of the genomic database, the malicious SP will check   all SNPs at the same loci of   %parents and the children, 
family members and then flips SNPs at the  loci that violate   Mendel's law. For example, if both parents have SNP value 0, but their children have SNP value 1 at the same locus, then, the malicious SP knows that one or more family members' SNP values have been changed with high probability due to fingerprint insertion, since such   change can   be due to a mutation with a slight probability. 
% In row-wise correlation attack,   i.e., $\mathrm{Atk_{row}}(\mathcal{S})$, we assume that the database recipients (i.e., SPs) know the existence of families and their members by checking their IDs (i.e., the primary keys of each individual, which is required to be immutable   in DBMS (Database Management System) design) and  the correlations (denoted as $\mathcal{S}$) among family members (measured in terms of similarity, e.g., the inner product between the SNP sequences of parents and their children). 
Next, the malicious SP can further  calculates the empirical row-wise correlations (denoted as $\mathcal{S}'$) from the received fingerprinted copy,   compares $\mathcal{S}'$ with   $\mathcal{S}$, and   changes   entries that leads to large discrepancy between   them.

\noindent\textbf{Column-wise correlation attack $\mathrm{Atk_{col}}(\mathcal{J})$.} We model the publicly known 
allelic associations (linkage disequilibrium) between SNP values at different loci as a set of joint distributions $\mathcal{J}$.  %The malicious SP has prior knowledge about the correlations among each pair of SNPs  characterized by the set of joint probability distributions $\mathcal{J}$. 
Once a malicious SP receives  the fingerprinted    database, it   can calculate a new set of  empirical pair-wise joint distributions $\mathcal{J}'$. 
Then, it compares $\mathcal{J}'$ and  $\mathcal{J}$, and flips the entries in the fingerprinted copy that leads to large discrepancy between them.

% \hl{add this: For the genomic database we consider, columns correspond to point mutations (SNPs) and rows correspond to individuals.  
% After launching these attacks on a fingerprinted database, the malicious SP can easily distort the added fingerprint to mislead the fingerprint extraction algorithm and cause the database owner to accuse innocent parties.}

% \end{itemize}

In this paper we do not consider other common attacks, such as the  subset attack and superset attack ~\cite{li2005fingerprinting},  because they are usually much weaker than the random bit flipping attack  as shown in \cite{yilmaz2020collusion}. Another widely investigated attack is the  collusion attack \cite{boneh1998collusion,boneh1995collusion,pfitzmann1997asymmetric}. Our proposed robust fingerprinting scheme for genomic databases can also be extended to  collusion-resistant  genomic database fingerprinting by incorporating collusion-resistant codewords when generating the fingerprint \cite{boneh1995collusion}. We will extend our work in the scenario of colluding medical SPs in future. 

\subsection{Fingerprint Robustness Metrics}
\label{sec:robustness_metrics}
%\vspace{-1mm}
The primary goal of a malicious SP is to distort the fingerprint in %$\widetilde{\mathbf{R}}$, 
its received copy   to avoid  being accused. We use the percentage  of compromised fingerprint bits, i.e.,  $\mathrm{Per_{cmp}}$, to measure the robustness of the fingerprint scheme. 
%In this paper, we  focus on the following fingerprint  robustness metric. % about a pirated  database  $\overline{\mathbf{R}}$ generated by launching attacks on $\widetilde{\mathbf{R}}$.
% \begin{itemize}
% \noindent\textbf{Percentage  of compromised fingerprint bits, $\mathrm{Per_{cmp}}$.} 
$\mathrm{Per_{cmp}}$ calculates  the percentage  of mismatches between the fingerprint  bit-sting extracted from the compromised fingerprinted database and the original fingerprint bit-string that is used to generate the fingerprinted genomic database. In our previous work \cite{ji_fp_raid2021}, we have shown that if  the malicious SP can compromise more than 50\% of the fingerprint bits, it can cause the database owner to accuse other  innocent SPs who also received the databases. In this paper, we  only focus on $\mathrm{Per_{cmp}}$, because other robustness metrics (e.g., the accusable ranking of a malicious SP) directly depends on $\mathrm{Per_{cmp}}$ \cite{ji_fp_raid2021}.

\subsection{Utility Metrics}
\label{sec:utility_metrics}
%\vspace{-1mm}
% \hl{EA: also need to mention the attacker's utility (and highlight that the attacker can succeed easily by over-distorting. But, a rational attacker does not want to over-distort  due to its utility requirement)}

Fingerprinting naturally changes the content of   databases (i.e., the values of the SNPs), and hence degrades its utility. 
We quantify the utility of a fingerprinted genomic database using the following metrics. 
% \begin{itemize}

\noindent\textbf{Accuracy of the database, i.e., $Acc$.} %$\widetilde{\mathbf{R}}$,} %We quantify the accuracy of $\widetilde{\mathbf{R}}$ as 
It calculates the percentage of matched data entries between the original genomic database and the fingerprinted copy (or the compromised fingerprinted copy, i.e., the pirated copy generated by a malicious SP). The higher $Acc$, the fewer entries are changed due to fingerprint insertion, attack from the malicious SP, or mitigation to resist the attacks, and thus, the higher the utility.
% i.e., $Acc(\widetilde{\mathbf{R}}) = 1- \widetilde{\mathbf{R}}\oplus \mathbf{R}/(M*L)$, 
% % \begin{equation*}
% %       Acc(\widetilde{\mathbf{R}}) = 1- \widetilde{\mathbf{R}}\oplus \mathbf{R}/(M*L),
% %   \end{equation*}
% where $\oplus$ is the symmetric difference operator that counts   the number of different entries in the fingerprinted and the original databases. $Acc(\widetilde{\mathbf{R}})$ measures the percentage of matched entries between the    fingerprinted and the original databases. % (higher $Acc(\widetilde{\mathbf{R}})$ means higher utility for $\widetilde{\mathbf{R}}$). 

% \hl{add p-value}  

\noindent\textbf{Consistency of SNP-phenotype association.} GWAS (genome-wide association study)  is a widely adopted method   to identify genetic variations that are
associated with a particular phenotype (e.g.,  a disease). In   GWAS, a researcher usually  quantifies the associations between
a   phenotype  and each SNP in the database using a
 p-value with a 
confidence level of 95\% \cite{sheskin2003inferential,halimi2021privacy}. In particular, SNPs with low p-values   (typically
smaller than 0.05) are considered to have strong associations with the phenotype (i.e., the association cannot be due to chance). In general, a larger utility loss in terms of accuracy degradation will lead to less accurate SNP-phenotype association.  To evaluate the p-value of each SNP in the genomic database, we first randomly divide  the database into non-overlapping case (denoted as $S$) and control (denoted as $C$) groups, and then follow the   steps listed in (\ref{eq:p-value}) to perform the calculations.

% \begin{wrapfigure}{l}{.3\textwidth}
% %\vspace{-6mm}
% \begin{tiny}
\begin{equation}
\begin{aligned}
   & OR  =  \frac{C_0(S_1+S_2)}{S_0(C_1+C_2)}, \\
    &StdErr(\ln(OR))\\
   & =  \sqrt{\frac{1}{S_1+S_2}+\frac{1}{S_0}+\frac{1}{C_1+C_2}+\frac{1}{C_0}},\\
   & z  = \frac{\ln(OR)}{StdErr(\ln(OR))},\\
   & p =  \Psi(-z)+1-\Psi(z).
    \end{aligned}
    \label{eq:p-value}
\end{equation}
% \end{tiny}
% %\vspace{-6mm}
% \end{wrapfigure}
% \noindent 

In particular, in (\ref{eq:p-value}), $OR$ is   the odd ratio, $C_0$, $C_1$, and $C_2$ (or $S_0$, $S_1$, and $S_2$) are the numbers representing a specific SNP taking a value of 0, 1, and 2 in the control (or case) group. $StdErr(\ln(OR))$ is the standard error of the logarithm of the   odd ratio, and $z$ is interpreted as the   standard normal deviation (i.e., $z$-value). Finally, the p-value is the   area (probability) of the normal distribution that falls outside $\pm z$, and it can be obtained using $\Psi(\cdot)$; the cumulative distribution function   of the standard normal distribution.

To evaluate the utility of the genomic database, we identify the top-50 SNPs (i.e., the  50 SNPs with the lowest p-values) from the original (non-fingerprinted) database. Then, we check how many of such SNPs are preserved (i.e., remains to be the top-50 SNPs)  after fingerprinting or various attacks.  Note that we only consider the consistency of SNP-phenotype association for individual SNPs (i.e., not the consistency of SNP-phenotype association for SNP tuples). This is because the proposed mitigation techniques can preserve the Pearson’s correlations among each pair of SNPs (see Section \ref{sec:col_defense} for details). 
\section{The Foundation:  Vanilla Genomic Fingerprinting Scheme}\label{sec:vanilla_fp}

Now, we establish the foundation of our robust fingerprinting scheme for genomic database. Our developed vanilla scheme is inspired by \cite{li2005fingerprinting}, which enables  the insertion and extraction of arbitrary bit-strings in  databases. However, our   scheme differs from \cite{li2005fingerprinting} and its   variants, e.g.,    \cite{guo2006fingerprinting,ji_fp_raid2021}, as they only mark one bit position in each selected row, which leads to less fingerprinting robustness due to significantly larger number of attributes (e.g., number of SNPs in a genomic database) and strong correlation patterns  in genomic data. 

%\vspace{-3mm}
\subsection{Fingerprint Insertion Phase of the Vanilla Scheme}\label{sec:vanilla_insert}

When the database owner  shares a fingerprinted copy of the genomic database $\mathbf{R}$ with a SP (whose id is $n$), it first generates the fingerprint bit string $\mathbf{f}_{\mathrm{SP}_n}=Hash(\mathcal{K}|n)$, where $\mathcal{K}$ is the secret key of the database owner and $|$ stands for the concatenation operator.\footnote{In this paper, we use MD5 to generate a 128-bits fingerprint string, because if the database owner shares $C$ copies of its database, then as long as $L\geq \ln C$, the fingerprinting mechanism can thwart exhaustive 
search and various types of attacks, and in most cases a 64-bits
fingerprint string is shown to provide high robustness \cite{li2005fingerprinting}.} The database owner uses    a cryptographic pseudorandom sequence generator $\mathcal{U}$ to select   specific bit positions of specific SNPs from some individuals and fingerprint these bits using  mark bits $m$'s, which are the result of the XOR operation between the random mask bits ($x$'s) and   randomly selected fingerprint bits ($f_l$'s), i.e., $m = x\oplus f_l$, where $f_l$ is the $l$th bit in $\mathbf{f}_{\mathrm{SP}_n}$.

To be more specific, for all individuals in the genomic database, the database owner fingerprints the SNP sequence if $\mathcal{U}_1(\mathcal{K}|\boldsymbol{r}_i.\mathrm{primary\ key})\ \mathrm{mod}\ \floor{1/\gamma_r} = 0$, where $\gamma_r\in(0,1)$ is the row fingerprint density. As a result, the fraction of fingerprinted SNP sequences  in $\mathbf{R}$ is 
approximately $\gamma_r$. For all SNPs in a selected sequence (i.e., $\boldsymbol{r}_i$), the element with attribute $p$  (i.e., the SNP value at loci $p$ of $\boldsymbol{r}_i$ represented by $\boldsymbol{r}_i[p]$) will be fingerprinted if $\mathcal{U}_2(\mathcal{K}|\boldsymbol{r}_i.\mathrm{primary\ key}| p)\ \mathrm{mod}\ \floor{1/\gamma_l} = 0$, where $\gamma_l\in(0,1)$ is the column fingerprint density. Then, the database owner sets the binary mask bit $x= \mathcal{U}_3(\mathcal{K}|\boldsymbol{r}_i.\mathrm{primary\ key}| p)\ \mathrm{mod}\ 2$, and selects one bit position of $\mathbf{f}_{\mathrm{SP}_n}$ via $l=\Big(\mathcal{U}_4(\mathcal{K}|\boldsymbol{r}_i.\mathrm{primary\ key}|p)\ \mathrm{mod}\ L\Big)+1$. Next, it obtains the mark bit $m$ as $m = x\oplus \mathbf{f}_{\mathrm{SP}_n}[l]$, and selects a bit position (count backwards) of $\boldsymbol{r}_i[p]$ via   $t = \Big(\mathcal{U}_5(\mathcal{K}|\boldsymbol{r}_i.\mathrm{primary\ key}|p)\ \mathrm{mod}\ 2\Big)+1$. Finally, it fingerprints $\boldsymbol{r}_i[p]$ by replacing the $t$th to the last bit of $\boldsymbol{r}_i[p]$ with $m$. We summarize the steps of the fingerprint insertion phase of the vanilla fingerprinting scheme in Algorithm \ref{algo:vanilla-insert}.
%\vspace{-6mm}

% \setlength{\textfloatsep}{0.1cm}
\begin{algorithm}
% \scriptsize
\small
\SetKwInOut{Input}{Input}
\SetKwInOut{Output}{Output}
\Input{The original genomic relational database $\mathbf{R}$, row fingerprinting density $\gamma_r$, column fingerprinting density $\gamma_l$, database owner's secret key $\mathcal{K}$, pseudorandom number sequence generator $\mathcal{U}$, and the SP's series number $n$ (which can be public).}
\Output{The vanilla fingerprinted genomic relational database $\R$.}

Generate the fingerprint bit string of SP $n$, i.e., $\mathbf{f}_{\mathrm{SP}_n} = Hash(\mathcal{K}|n)$;

\ForAll{\text{individual} $\boldsymbol{r}_i\in\mathbf{R}$}{
\If{$\mathcal{U}_1(\mathcal{K}|\boldsymbol{r}_i.\mathrm{primary\ key})\ \mathrm{mod}\ \floor{1/\gamma_r} = 0$}{
\CommentSty{{\color{gray}//fingerprint the SNP sequence of the $i$th individual}}

\ForAll{SNP element  $\boldsymbol{r}_i[p]\in\boldsymbol{r}_i$}{

\If{$\mathcal{U}_2(\mathcal{K}|\boldsymbol{r}_i.\mathrm{primary\ key}| p)\ \mathrm{mod}\ \floor{1/\gamma_l} = 0$}{
\CommentSty{{\color{gray}//fingerprint the $p$th  SNP      of the $i$th individual}}

% $\mathrm{attribute\_index}\ p = \mathcal{U}_2(\mathcal{K}|\boldsymbol{r}_i.\mathrm{primary\ key})\ \mathrm{mod}\ |\mathcal{F}|$.
% \CommentSty{//fingerprint this attribute}

Set $\mathrm{mask\_bit}\ x = 0$, if $\mathcal{U}_3(\mathcal{K}|\boldsymbol{r}_i.\mathrm{primary\ key}|p)$ is even; otherwise set $x=1$.

$\mathrm{fingerprint\_index}\ l=\mathcal{U}_4(\mathcal{K}|\boldsymbol{r}_i.\mathrm{primary\ key}|p)\ \mathrm{mod}\ L$.

$\mathrm{fingerprint\_bit}\ f_l = \mathbf{f}_{\mathrm{SP}_n}(l)$.

% \CommentSty{//select fingerprint bit position $l$}

$\mathrm{mark\_bit}\ m = x\oplus f_l$.

Set $t = \Big(\mathcal{U}_5(\mathcal{K}|\boldsymbol{r}_i.\mathrm{primary\ key}|p)\ \mathrm{mod}\ 2\Big)+1$.

Set the $t$th to the last bit of $\boldsymbol{r}_i[p]$ to $m$.

}

}

} 

}

%  Return $\widetilde{\mathbf{R}}\Big(\mathrm{FP},\emptyset,\emptyset \Big)$.

		\caption{Vanilla   scheme: fingerprint insertion}
\label{algo:vanilla-insert}
\end{algorithm}
% \setlength{\floatsep}{0.1cm}
%\vspace{-4mm}

% \hl{highlight the differences with existing vanillas, why we modify the 1st vanilla, robust fp is independent on the adopted vanilla, which can be determined by the required utility metrics.}

% It is noteworthy that our proposed vanilla fingerprinting scheme differs from the ones considered in \cite{li2005fingerprinting,guo2006fingerprinting,ji_fp_raid2021}, because these schemes only mark one bit position in each selected row, which leads to less fingerprinting robustness due to significantly larger number of attributes in genomic data (e.g., number of SNPs in a genomic database).  

% In contrast, 

Different from \cite{li2005fingerprinting,guo2006fingerprinting,ji_fp_raid2021}, by involving $\gamma_l$, % (the column fingerprinting density), 
our vanilla scheme can mark more bits in each selected row. In our recent work \cite{ji2021differentially}, we also derived a closed-form expression to characterize the relationship between fingerprint robustness and database utility. Thus, by jointly tuning $\gamma_r$ and $\gamma_l$ we can also achieve desired tradeoff between robustness and utility. We will theoretically and empirically investigate this in the future.

\subsection{Fingerprint Extraction Phase of the Vanilla Scheme}\label{sec:vanilla_extract}

When the database owner observes a leaked (or pirated) genomic database denoted as $\overline{\mathbf{R}}$, it tries to identify the traitor (i.e.,  the malicious SP) by extracting the fingerprint from $\overline{\mathbf{R}}$ and comparing it with the fingerprints of all SPs who have received its genomic database.

We present the fingerprint extraction phase of the vanilla scheme in Algorithm \ref{algo:vanilla_extract}. Specifically, the database owner first initiates a fingerprint template $\mathbf{f} = (f_1,f_2,\cdots,f_L) = (?,?,\cdots,?)$.  Here, ``$?$'' means that the fingerprint bit at that position remains to be determined.\footnote{Similar symbol has also been used in other works   \cite{boneh1995collusion,li2005fingerprinting,agrawal2003watermarking,ji_fp_raid2021}.} Then, the database owner determines the mask bit ($x$), obtains the corresponding    indices of   fingerprint bits ($l$'s),   locates the bit positions of the fingerprinted SNP elements exactly as in the fingerprint insertion phase, and finally fills in each ``$?$'' using a voting scheme.  To be more precise, for each fingerprinted SNP %(potentially compromised, i.e., distorted by the attacker) 
$\overline{\boldsymbol{r}_i}[p]$, the database owner obtains the corresponding mark bit $m$ by reading the $t$th to the last bit of  $\overline{\boldsymbol{r}_i}[p]$, and recovers one instance of the $l$th bit of the fingerprint bit string via $f_l = m\oplus x$.  Since the value of $f_l$ may be changed by the malicious SP, the database owner maintains and updates two counting arrays $\mathbf{c}_0$ and $\mathbf{c}_1$, where $\mathbf{c}_0(l)$ and $\mathbf{c}_1(l)$ record the number of times $f_l$ is recovered as 0 and 1, respectively.  Finally, the database owner sets $\mathbf{f}(l) = 1$ if $\mathbf{c}_1(l)/(\mathbf{c}_1(l)+\mathbf{c}_0(l))\geq \tau$, and $\mathbf{f}(l) = 0$ if $\mathbf{c}_0(l)/(\mathbf{c}_1(l)+\mathbf{c}_0(l))\geq \tau$, otherwise $\mathbf{f}(l) = ?$ (i.e., this fingerprint bit cannot be determined due to the database owner's low confidence), where $\tau\in(0.5,1]$ is the parameter that quantifies the database owner's confidence in the fingerprint recovery phase.\footnote{In this paper, we set $\tau=0.7$, which implies the database owner has higher confidence during fingerprint extraction than the other works, e.g., in \cite{ji_fp_raid2021,li2005fingerprinting} $\tau$ is slightly higher than 0.5.}
%\vspace{-6mm}

\begin{algorithm}
% \scriptsize
\small
\SetKwInOut{Input}{Input}
\SetKwInOut{Output}{Output}
\Input{The leaked genomic database $\overline{\mathbf{R}}$, row fingerprinting density $\gamma_r$, column fingerprinting density $\gamma_l$, database owner's secret key $\mathcal{K}$, pseudorandom number sequence generator $\mathcal{U}$,   and a fingerprint template $(?,?,\cdots,?)$, where $?$ represents unknown value.}
\Output{The extracted fingerprint bit string $\mathbf{f}$ from the leaked database.}

Initialize $\mathbf{c}_0(l) = \mathbf{c}_1(l)=0, \forall l\in[1,L]$.

% Generate the fingerprint bit string of SP $n$, i.e., $\mathbf{f}_{\mathrm{SP}_n} = Hash(\mathcal{K}|n)$;

% \CommentSty{//scan all data records and obtain the counts for each fingerprint bit}

\ForAll{\text{individual} $\overline{\boldsymbol{r}_i}\in\overline{\mathbf{R}}$}{
\If{$\mathcal{U}_1(\mathcal{K}|\overline{\boldsymbol{r}_i}.\mathrm{primary\ key})\ \mathrm{mod}\ \floor{1/\gamma_r} = 0$}{
% \CommentSty{{\color{gray}//fingerprint the SNP sequence of the $i$th individual}}

\ForAll{SNP element  $\overline{\boldsymbol{r}_i}[p]\in\boldsymbol{r}_i$}{

\If{$\mathcal{U}_2(\mathcal{K}|\overline{\boldsymbol{r}_i}.\mathrm{primary\ key}| p)\ \mathrm{mod}\ \floor{1/\gamma_l} = 0$}{
% \CommentSty{{\color{gray}//fingerprint the $p$th  SNP      of the $i$th individual}}

% $\mathrm{attribute\_index}\ p = \mathcal{U}_2(\mathcal{K}|\boldsymbol{r}_i.\mathrm{primary\ key})\ \mathrm{mod}\ |\mathcal{F}|$.
% \CommentSty{//fingerprint this attribute}

Set $\mathrm{mask\_bit}\ x = 0$, if $\mathcal{U}_3(\mathcal{K}|\overline{\boldsymbol{r}_i}.\mathrm{primary\ key}|p)$ is even; otherwise set $x=1$.

$\mathrm{fingerprint\_index}\ l=\mathcal{U}_4(\mathcal{K}|\overline{\boldsymbol{r}_i}.\mathrm{primary\ key}|p)\ \mathrm{mod}\ L$.

% $\mathrm{fingerprint\_bit}\ f = \mathbf{f}_{\mathrm{SP}_n}(l)$.

% \CommentSty{//select fingerprint bit position $l$}

% $\mathrm{mark\_bit}\ m = x\oplus f$.

Set $t = \Big(\mathcal{U}_5(\mathcal{K}|\overline{\boldsymbol{r}_i}.\mathrm{primary\ key}|p)\ \mathrm{mod}\ 2\Big)+1$.

Set the $\mathrm{mark\_bit}$ $m$  as the $t$th to the last bit of $\overline{\boldsymbol{r}_i}[p]$.

% Set the $t$th to the last bit of $\overline{\boldsymbol{r}_i}[p]$ to $m$.

Recover the fingerprint bit $f_l = m\oplus x$.

$\mathbf{c}_1(l)++$, if $f_l=1$; otherwise $\mathbf{c}_0(l)++$.

}

}

} 

}

\ForAll{$l\in[1,L]$}{
% \If{count[l][0] = count[l][1]}{return none suspected}

% $\mathbf{c}_1(l)/(\mathbf{c}_1(l)+\mathbf{c}_0(l))\geq \tau$, and $\mathbf{f}(l) = 0$ if $\mathbf{c}_0(l)/(\mathbf{c}_1(l)+\mathbf{c}_0(l))\geq \tau$

$\mathbf{f}(l)=1$, if $\mathbf{c}_1(l)/(\mathbf{c}_1(l)+\mathbf{c}_0(l))\geq \tau$, and $\mathbf{f}(l) = 0$,  if $\mathbf{c}_0(l)/(\mathbf{c}_1(l)+\mathbf{c}_0(l))\geq \tau$
.\label{line:majority} 
}

%  Return the extracted fingerprint bit string $\mathbf{f}$.

		\caption{Vanilla   scheme: fingerprint extraction}
\label{algo:vanilla_extract}
\end{algorithm}
%\vspace{-10mm}

%\vspace{-4mm}
\section{Consolidating the Foundation: Making the Vanilla Genomic Fingerprinting Scheme  Robust   Against   Correlation Attacks}
\label{sec:robust_fp}

Here, we propose a robust fingerprinting scheme for genomic databases against the correlation attacks identified in Section \ref{sec:threat_models}.  The robust scheme is developed by augmenting the vanilla scheme using two mitigation techniques which can serve as the post-processing steps  after the vanilla fingerprinting. It brings us two benefits: (i)   fingerprint robustness of the vanilla scheme is maintained, because the devised  mitigation techniques only change non-fingerprinted entries and (ii) the mitigation techniques can be applied to any vanilla  fingerprinting schemes to resist correlation attacks on genomic databases. As discussed in Section \ref{sec:vanilla_insert}, we choose our developed vanilla scheme to have more control over the fingerprint density on each selected SNP sequence. In practice, one can develop their own vanilla scheme depending on the content of their genomic database and then apply our proposed mitigation techniques to make their scheme also robust against the correlation attack.

%that can serve as post-processing steps for any off-the-shelf (vanilla) fingerprinting schemes.  
% \hl{EA: for J and S, I think we should use the input arguments (as in the notations table). It is important from which dataset they are computed.}
In highlevel, to provide robustness against the %Mendel's law-based correlation attack $\mathrm{Atk_{Mendel}}$,
row-wise correlation $\mathrm{Atk_{row}(\mathcal{S})}$ and column-wise correlation attack $\mathrm{Atk_{col}(\mathcal{J})}$, the database owner (Alice) will perform mitigation steps (after the vanilla fingerprinting) by utilizing Mendel's law and her prior knowledge $\mathcal{S}$ (correlation of genomic data among different individuals), $\mathcal{J}$ (correlation of SNP values at different loci) to reduce the discrepancy caused by fingerprinting insertion. 
 We will show that to implement the proposed mitigation steps, Alice needs to change only a few entries after the vanilla fingerprinting (e.g., less than $3\%$ as shown in Table \ref{table:chg_mtg} in Section \ref{sec:robust_exp}). % in $\widetilde{\mathbf{R}}(\mathrm{FP},\emptyset,\emptyset)$.
% \hl{EA: notation not consistent with the table, lets keep the input arguments fixed}, 
% such that the post-processed fingerprinted database has both  column-  and row-wise correlations close to $\mathcal{J}'$ and $\mathcal{S}'$.
% In practice,  $\mathcal{J}'$ and $\mathcal{S}'$ may be different from $\mathcal{J}$ and $\mathcal{S}$, which are used by   the malicious SP  to launch correlation attacks. For example, $\mathcal{J}'$ and $\mathcal{S}'$ can be directly calculated from the original database $\mathbf{R}$ \hl{EA: we said J and S are calculated directly from the dataset before, right? If so, this is confusing}, which is only available to the database owner, or they can be obtained by the database owner from a  resource that is different with the malicious SP. We refer to this as prior knowledge asymmetry between database owner and malicious SP, and we will empirically evaluate its impact on attacks and defenses in fingerprinted databases in Section \ref{.}

%\hl{EA: you can add something like "to the advantage of the malicious SP, we assume the malicious SP always has more accurate or at least equally accurate knowledge about column-wise joint distributions and row-wise statistical relationships}

% \subsection{Mitigating the Mendel's law-based correlation attack}\label{sec:mtg_mendel}

%\vspace{-4mm}
\subsection{Mitigating the Row-wise Correlation Attack}
\label{sec:row-defense}
%\vspace{-1mm}

To mitigate the row-wise correlation attack (in Section \ref{sec:threat_models}), we develop    $\mathrm{Mtg_{row}}(\mathcal{S})$, which is composed of two phases. The first phase tries to eliminate  all SNP loci that violate the Mendel's law, and the second phase makes the similarities of genome data among family members close to that before the fingerprint (generated by the vanilla scheme) is inserted.   In particular, for each family set denoted as $\mathrm{fmly}$, $\mathrm{Mtg_{row}}(\mathcal{S})$ 
% {\color{blue}We develop $\mathrm{Mtg_{Mendel}}$ to mitigate the Mendel's law-based correlation attack. The proposed mitigation technique 
checks all fingerprinted SNPs of all family members. 
%belonging to the members of a same family. 
%mother-father-child SNP tuples in each family set. 
%Assume the family members in the database include mother, father, and child. Then, the mitigation technique check all mother-father-child SNP tuples in the given family. 
If  the SNP-tuple at a locus violates  the Mendel's law,  $\mathrm{Mtg_{row}}(\mathcal{S})$ changes other non-fingerprinted entries in the tuple to make them comply with the Mendel's law. For example, if a mother-father-child SNP tuple at a specific locus takes value ``2-1-0'' (2 for mother, 1 for father, and 0 for child, which violates the Mendel's law), and ``1'' (value of the SNP for the mother) is the  fingerprinted SNP, then, $\mathrm{Mtg_{row}}(\mathcal{S})$ can modify this tuple as ``2-1-1''.

% \subsection{Mitigating the Row-wise Correlation Attack}
% \label{sec:row-defense}

% % \hl{EA: i didnt read this part, i think it is still incomplete}

% To mitigate the row-wise correlation attack (in Section \ref{sec:threat_models}), we develop    $\mathrm{Mtg_{row}}(\mathcal{S}')$.
% The main goal of $\mathrm{Mtg_{row}}(\mathcal{S}')$ is to avoid a malicious SP from distorting the fingerprint due to discrepancies in the   correlation of SNP sequences among family members.

In the second phase, for each family set (i.e., $\mathrm{fmly}$) in the genomic database,  %$\mathrm{Mtg_{row}}(\mathcal{S})$ 
Alice changes the SNP values of the family members 
% individuals belonging to the parent subset (i.e., $\mathrm{fmly_{prnt}}$) and the children subset (i.e., $\mathrm{fmly_{chldn}}$) 
such that the cumulative  similarities between individuals with kinship relation 
%parents and children 
is close to their original similarities.  %$\mathrm{Mtg_{row}}(\mathcal{S}')$ 
This can be formulated as the   following optimization problem
\begin{equation}
\begin{aligned}
\min_{\widetilde{\widetilde{\boldsymbol{r}_i}}} \quad &  
\sum_{i,j\in\mathrm{fmly}} \   \Big|{s_{ij}}^{\mathrm{fmly}}-\widetilde{\widetilde{s_{ij}}}^{\mathrm{fmly}}\Big|  \\
\textrm{s.t.} \quad & \widetilde{\widetilde{\boldsymbol{r}_i}} = \mathrm{value\  change}(\widetilde{\boldsymbol{r}_i}), \\%i \in\mathrm{fmly},\\
   & \widetilde{\widetilde{s_{ij}}}^{\mathrm{fmly}} = \langle\widetilde{\widetilde{\boldsymbol{r}_i}},\widetilde{\widetilde{\boldsymbol{r}_j}}\rangle. %i\in\mathrm{fmly_{prnt}},j \in\mathrm{fmly_{chldn}},
\end{aligned}
\label{eq:row_defense}
\end{equation}%\vspace{-11.5mm}
% \end{wrapfigure}

In (\ref{eq:row_defense}), ${s_{ij}}^{\mathrm{fmly}}$ stands for the publicly known  similarities between two family members  in $\mathrm{fmly}$ and $\widetilde{\widetilde{s_{ij}}}^{\mathrm{fmly}}$ stands for  the similarities after  row-wise  mitigation.\footnote{In this paper, the similarity between two individuals is defined as the inner product between their SNP sequences. One can also define the similarity using a  common metric adopted in biology, e.g., the allele shared distance (i.e., ASD) (page 308 \cite{park2013inference}), which is also related to the considered inner product.} We denote the SNP sequence of an individual $i$ after the mitigation   as $\widetilde{\widetilde{\boldsymbol{r}_i}}$. Also, $\mathrm{value\ change}(\cdot)$ is the function that changes each SNP attribute of $\widetilde{\boldsymbol{r}_i}$, and it will be elaborated later. 
(\ref{eq:row_defense})  is an integer programming problem. Since each SNP sequence may contain thousands of SNPs, it will be computationally expensive to obtain the optimal value of $\widetilde{\widetilde{\boldsymbol{r}_i}}$. Thus, we solve it heuristically. Particularly, if ${s_{ij}}^{\mathrm{fmly}}>\widetilde{s_{ij}}^{\mathrm{fmly}}$ ($\widetilde{s_{ij}}^{\mathrm{fmly}}$ is the empirical similarity calculated after the vanilla fingerprinting), the database owner needs to post-process  $\widetilde{\boldsymbol{r}_i}, \forall i \in\mathrm{fmly}$ to increase the similarity. On the other hand, if  ${s_{ij}}^{\mathrm{fmly}}<\widetilde{s_{ij}}^{\mathrm{fmly}}$, the database owner needs to post-process  $\widetilde{\boldsymbol{r}_i}, \forall i \in\mathrm{fmly}$ to decrease the similarity.

We use the  example of SNP sequences from mother-father-child to further explain how to increase or decrease the similarity. To increase $\widetilde{s_{ij}}^{\mathrm{fmly}}$, we randomly select a certain number of  non-fingerprinted 3-SNP-tuples (i.e., mother-father-child) with value ``0-0-0'' and then change it to ``1-0-1'' or ``0-1-1'' depending on whether $\widetilde{s_{ij}}^{\mathrm{fmly}}$ is a mother-child or father-child similarity. We only change 3-SNP-tuples with value ``0-0-0'', because this is the one of the most common 3-tuples in all families, and modification of mother-child tuples will not have an impact on the father-child similarity (vice-versa). To decrease  $\widetilde{s_{ij}}^{\mathrm{fmly}}$, we let the database owner change a certain number of 3-tuples with value ``1-0-1'' (or ``0-1-1'') to ``0-0-0'' if $\widetilde{s_{ij}}^{\mathrm{fmly}}$ is the mother-child (or father-child) similarity. The reasons are exactly the same as the case of   increasing  $\widetilde{s_{ij}}^{\mathrm{fmly}}$.     Although there are also higher degrees of relatedness among family members (e.g., the SNP  correlations between grandparents and grandchildren), those correlation (or similarity) are usually much weaker than the first order correlations (e.g., mother-father-child). The vanilla fingerprinting is subject to destroying the strong correlations the most (and hence an attacker can easily infer the fingerprint due to the distorted correlations). For higher degree family members, the original correlations are not high and fingerprint  will not destroy such correlation too much. We will experimentally verify this in Section \ref{sec:dataset}.

% {\color{red}Although the optimization problem in (\ref{eq:row_defense}) is a generic approach to mitigate the row-wise correlation attack, we }

Note that the row-wise mitigation technique for genomic databases is different than the one developed for general databases in our previous work \cite{ji_fp_raid2021}, which changes entries of non-fingerprinted data records to  make the newly obtained  similarities as far away from Alice's prior knowledge $\mathcal{S}$ to mislead the malicious SP. In contrast, here, we make the new similarities  close to $\mathcal{S}$ in order to alleviate   $\mathrm{Atk_{row}}(\mathcal{S})$ (try to make $\mathrm{Atk_{row}}(\mathcal{S})$ distort less fingerprint bits), and if the objective function on (\ref{eq:row_defense}) equals 0, $\mathrm{Atk_{row}}(\mathcal{S})$ is completely invalidated. The reason that we can pursue this in genomic database is because each row has much more attributes then general databases and the number of unique values is small  (i.e., only 3 options: 0, 1, and 2). 
Besides, the row-wise mitigation techniques developed in \cite{ji_fp_raid2021}  solves an NP-hard  combinatorial  search  problem greedily, which introduces large computation overhead.

%\vspace{-4mm}
\subsection{Mitigating Column-wise Correlation Attack}
\label{sec:col_defense}
%\vspace{-1mm}
% \hl{EA: as mentioned before, I think when you dont write the input arguments for J, J', etc, it becomes confusing. Based on the notation table both J and J' can be denoted in the same way with different datasets as the input arguments; for J, input argument can be R, for J', input argument can be a noisy version of R.}

% \subsubsection{Mitigation via mass  transport} 
To make the  vanilla %genomic database fingerprinting 
scheme robust against column-wise correlation attack, we propose  $\mathrm{Mtg_{col}}(\mathcal{J})$, which transforms the vanilla fingerprinted genomic database   to have column-wise joint distributions (e.g., linkage disequilibrium between the SNPs) close to  the publicly known joint distributions in $\mathcal{J}$. Inspired by \cite{ji_fp_raid2021}, we develop   $\mathrm{Mtg_{col}}(\mathcal{J})$ using the idea of ``optimal  transport''~\cite{cuturi2013sinkhorn}, which moves the probability  mass of the marginal distribution of each SNP attribute of the vanilla fingerprinted genomic database  to resemble the distribution obtained from the marginalization of each reference joint distribution in $\mathcal{J}$. Then, the optimal  transport plan is used to change the entries in %each attribute of 
the genomic database after the vanilla fingerprinting. % $\widetilde{\mathbf{R}}(\mathrm{FP},\emptyset,\emptyset)$ to obtain $\widetilde{\mathbf{R}}\big(\mathrm{FP},\emptyset,\mathrm{Mtg_{col}}(\mathcal{J}')\big)$. 

In particular, for a specific SNP (column, i.e., locus of SNP sequence) $p$, we denote its marginal distribution obtained after the vanilla fingerprinting as $\Pr(C_{\widetilde{p}})$, and that obtained from the  marginalization of a joint distribution $J_{p,q}$ distribution in $\mathcal{J}$ as $\Pr(C_{p}) = J_{p,q}\mathbf{1}^T$  (here $q$ can be any attribute that is different from $p$, because the marginalization with respect to $p$ using different $J_{p,q}$ will lead to the identical marginal distribution of $p$).
% \hl{EA: we can probably obtain many different marginal distribution for p using marginalization of different joint distributions. Which one do we end up using?}. 
To move the mass of $\Pr(C_{\widetilde{p}})$ to resemble $\Pr(C_{p})$, we need to find another joint distribution (i.e., the mass  transport plan)  $G_{\widetilde{p},p}\in\R^{3\times 3}$, whose marginal distributions are identical to $\Pr(C_{\widetilde{p}})$ and $\Pr(C_{p})$. 
% \hl{EA: the way we use a and b below is not very clear to me}
%Let $a$ and $b$ be two distinct values that attribute $p$ can take ($a,b\in[0,k_p-1]$). 
Then,  $G_{\widetilde{p},p}(a,b), a,b\in\{0,1,2\}$ indicates that the database owner should change $G_{\widetilde{p},p}(a,b)$ percentage of entries in the vanilla fingerprinted genomic database   %$\widetilde{\mathbf{R}}(\mathrm{FP},\emptyset,\emptyset)$
whose attribute $p$ (SNP $p$) takes value $a$ %(i.e., $p=a$) 
to value $b$, % (i.e., change them to make $p=b$), 
so as to  make $\Pr(C_{\widetilde{p}})$ close to $\Pr(C_{p})$. In practice, such a  transport plan can be obtained by solving a regularized optimal  transport problem, i.e., the entropy regularized Sinkhorn distance minimization  \cite{cuturi2013sinkhorn} as follows:  
% \begin{equation}
% \label{ot_defense}
% \begin{aligned}
% &d\Big(\Pr(C_{\widetilde{p}}),\Pr(C_{p'}),\lambda_p\Big) \\
% =& \min_{G_{\widetilde{p},p'}\in\mathcal{G}\big(\Pr(C_{\widetilde{p}}),\Pr(C_{p'})\big)}<G_{\widetilde{p},p'},\Theta_{\widetilde{p},p'}>_F-\frac{H(G_{\widetilde{p},p'})}{\lambda_{p}},
% \end{aligned}
% \end{equation}
% \setlength{\abovedisplayskip}{3pt}
% \setlength{\belowdisplayskip}{3pt}
% %\vspace{-2mm}
\begin{equation}
\label{ot_defense}
\begin{aligned}
&d\Big(\Pr(C_{\widetilde{p}}),\Pr(C_{p}),\lambda_p\Big) \\
=& \min_{G_{\widetilde{p},p}\in\mathcal{G}\big(\Pr(C_{\widetilde{p}}),\Pr(C_{p})\big)}\langle G_{\widetilde{p},p},\Theta_{\widetilde{p},p}\rangle_F-\frac{H(G_{\widetilde{p},p})}{\lambda_{p}},
\end{aligned}
\end{equation}
% %\vspace{-2mm}
%\vspace{-3mm}
% \begin{equation}
% \label{ot_defense}
% d\Big(\Pr(C_{\widetilde{p}}),\Pr(C_{p}),\lambda_p\Big) 
% = \min_{G_{\widetilde{p},p}}\langle G_{\widetilde{p},p},\Theta_{\widetilde{p},p}\rangle_F-\frac{H(G_{\widetilde{p},p})}{\lambda_{p}},
% %\vspace{-3mm}
% \end{equation}
where   $G_{\widetilde{p},p}\in\mathcal{G}\big(\Pr(C_{\widetilde{p}}),\Pr(C_{p})\big) = \big\{G\in\R^{k_p\times k_p}\big|G\mathbf{1} = \Pr(C_{\widetilde{p}}),G^T\mathbf{1}  = \Pr(C_{p})\big\}$ is the set of all  joint probability distributions whose marginal distributions are the probability mass functions   of  $\Pr(C_{\widetilde{p}})$ and $\Pr(C_{p})$. $\langle\cdot,\cdot\rangle_F$ denotes the Frobenius inner product 
% (summation of     element-wise products) 
of two matrices. Also, $\Theta_{\widetilde{p},p}$ is the  transport cost matrix and $\Theta_{\widetilde{p},p}(a,b)>0$ representing the cost to move a unit percentage of mass from $\Pr(C_{\widetilde{p}} = a)$ to $\Pr(C_{\widetilde{p}} = b)$.  Finally, $H(G_{\widetilde{p},p}) = -\langle G_{\widetilde{p},p},\log G_{\widetilde{p},p}\rangle_F$ calculates  the information entropy of $G_{\widetilde{p},p}$ and $\lambda_p>0$ is a tuning parameter. In practice, (\ref{ot_defense}) can be efficiently solved by linear programming \cite{cuturi2013sinkhorn}. 
% iteratively rescaling rows and columns of the initialized $G_{\widetilde{p},p'}$ to have desired marginal distributions.  
The obtained $G_{\widetilde{p},p}$ is more heterogeneous for larger values of $\lambda_p$, i.e., the database owner will   change less entries after the vanilla fingerprinting, which preserves more database utility. On the contrary,   $G_{\widetilde{p},p}$ is more homogeneous for smaller values of $\lambda_p$, i.e., it causes more SNP entries to be changed, which leads to more utility loss.   Although (\ref{ot_defense}) processes each column (attribute) of the genomic database independently, as shown in \cite{ji_fp_raid2021}, the post-processed fingerprinted database will have the    Pearson’s
correlations among attribute pairs that are close to the prior knowledge $\mathcal{J}$. This further suggests that the mitigation step can boost the utility of the fingerprinted genomic databases.

\noindent\textbf{Mitigation against additional auxiliary information.} The developed row- and column-wise mitigation techniques focus on the correlation attacks that use generic correlations among genome data. In some task-dependent applications, the malicious SP can also use specific  auxiliary information, e.g., race-specific information determined by  genome and population structure, to compromise the fingerprinted database. This can be alleviated by involving additional mitigation steps before $\mathrm{Mtg_{row}}(\mathcal{S})$ and  $\mathrm{Mtg_{col}}(\mathcal{J})$ to make the vanilla fingerprinted database also match  those auxiliary information. More discussion on the availability of these information to the database owner is deferred to Section \ref{sec:discussion}.

\section{Experiment results}
\label{sec:geo_experiment}
%\vspace{-1mm}
In this section, we first  show that the   vanilla     fingerprinting  scheme can resist  random bit flipping attacks, but it is vulnerable to the correlation attacks developed specific for genomic databases. The correlation attacks are more powerful, as they can easily distort more than half of the embedded fingerprint bits at only a small cost of database utility (i.e., introducing less error and preserving the SNP-phenotype association). Then, we demonstrate that the proposed mitigation techniques can thwart correlation attacks and make the vanilla scheme   robust against them. Since the mitigation techniques only change limited entries on top of the vanilla scheme, they also maintain a high utility for the genomic database. More importantly, we show that if the attacker conducts correlation attacks after the proposed robust fingerprinting scheme, it cannot succeed even if at a significant cost of database utility loss.  Similar to \cite{ji_fp_raid2021}, since
 the row-wise correlation attack and mitigation are computationally light and modify less database entries, we let the malicious SP launches $\mathrm{Atk_{row}}(\mathcal{S})$ followed by $\mathrm{Atk_{col}}(\mathcal{J})$ when compromising a fingerprinted database, and let  the database owner perform $\mathrm{Mtg_{row}}(\mathcal{S})$ followed by $\mathrm{Mtg_{col}}(\mathcal{J})$ when making a vanilla fingerprinted database robust.

%\vspace{-4mm}
\subsection{Genomic Database Description}\label{sec:dataset}

\begin{figure}%{R}{0.45\linewidth}
\centering
%\vspace{-7mm}
\includegraphics[width=0.9\linewidth]{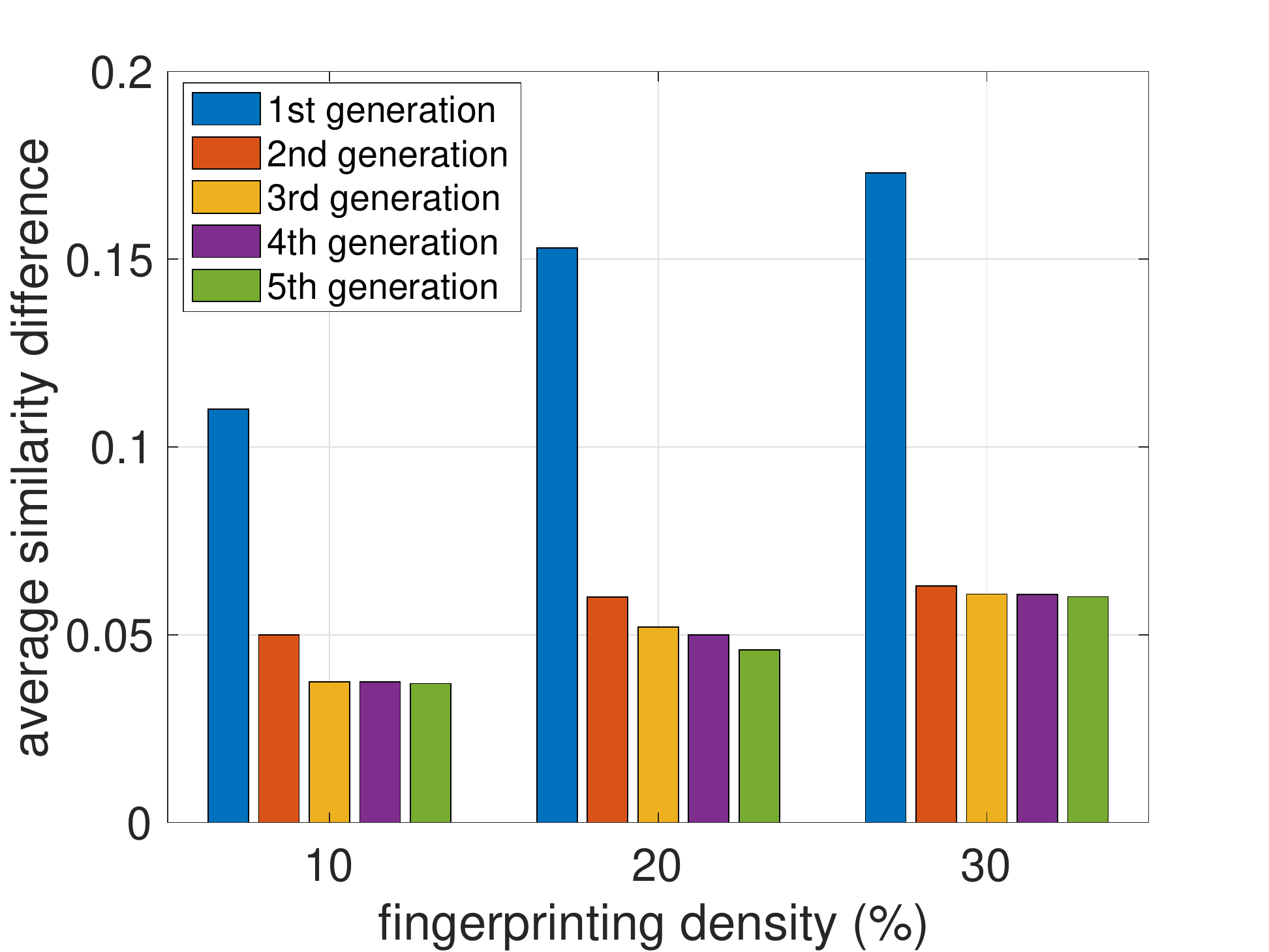}
%\vspace{-6mm}
\caption{\label{fig:generation_diff}Average absolute value of the SNP cosine similarity difference, before and after fingerprint insertion, among family members  and their different  generations of simulated descendants.}
%\vspace{-6mm}
\end{figure}

We use the SNP data belonging to 1500 individuals from the HapMap dataset~\cite{gibbs2003international}.  
Each individual has $156$ data points (i.e., SNPs). In this population, there are 150 families, each of which is composed of 3 individuals, i.e., mother, father, and child.  We assume that both the database owner  and the malicious SP know the members of each family and the pairwise correlations (e.g., linkage disequilibrium) among SNPs (Section \ref{sec:threat_models} discusses why this assumption is valid in practice).

\noindent\textbf{The importance to consider the correlations due to the first degree relationships among family members.}  As discussed in Section \ref{sec:row-defense}, row-wise correlations are the strongest among the first degree family members and  the vanilla fingerprinting potentially destroys such strong correlations the most (compared to the correlations between more distant family members). 
Here, we verify this claim by generating new generations of family members from the offsprings of the 150 families and checking the SNP similarities  between these new generations and the original 150 pairs of parents before and after fingerprint insertion. 
In Figure \ref{fig:generation_diff}, by varying the overall fingerprinting density ($\gamma_r\gamma_l\in\{10\%,20\%,30\%\}$), we plot the average absolute  difference of cosine similarity between the 150 parents and various generations of their descendants due to the vanilla fingerprinting. Clearly, the average change in the similarity with the first generation is the most significant for all considered fingerprinting density, which suggests that mother-father-child trio has the strongest correlation and it provides the most prior knowledge for the malicious SP to launch the row-wise correlation attack, and hence the row-wise mitigation techniques should preserve the correlations between first degree family members as much as possible. 
Thus, in the following experiment, we focus on the similarity change between individuals and their first generation of descendants during  row-wise correlation attack and mitigation.

\subsection{Vulnerability of Vanilla Fingerprinting Scheme Against Correlation Attack}\label{sec:vanilla_exp}
% In Table \ref{table:genome-all-attacks}, 
% in Appendix~\ref{sec:gen_result}, 

\subsubsection{Performance against Correlation Attack}

% \begin{wrapfigure}{R}{0.6\linewidth}
% \centering
% %\vspace{-7mm}
% \includegraphics[width=1\linewidth]{figures/performance_all_scheme.eps}
% %\vspace{-6mm}
%  \caption{Fingerprint robustness (i.e., percentage of compromised bits) versus utility loss (i.e., percentage of changed entries) when the vanilla scheme is compromised by the correlation attack (blue dots), the vanilla scheme is compromised by the random bit flipping  attack (black dots), and the robust scheme  is compromised by the correlation attack (red dots). Each dot represents a different experiment. Dots above the attack success boundary (green line) represent a successful attack (in which the database owner incorrectly blames an honest SP for the unauthorized sharing).}\label{fig:performance_all_scheme}
% %\vspace{-6mm}
% \end{wrapfigure}

\begin{figure}[htb]
% \begin{wrapfigure}{R}{0.58\linewidth}
   %\vspace{-10mm}
  \begin{center}
     \includegraphics[width=1\linewidth]{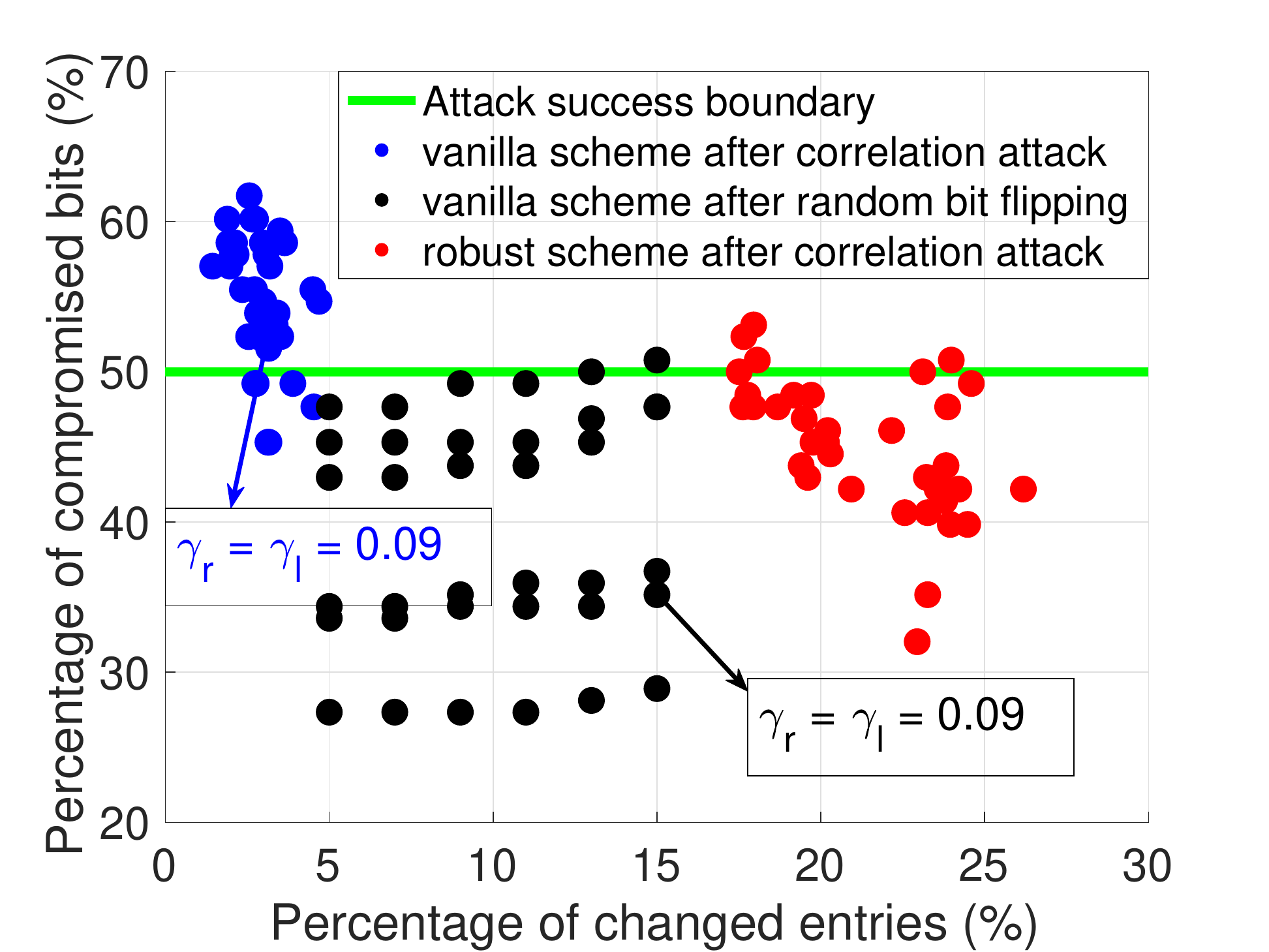}
      \end{center}
      %\vspace{-7mm}
  \caption{Fingerprint robustness (i.e., percentage of compromised bits) versus utility loss (i.e., percentage of changed entries) when the vanilla scheme is compromised by the correlation attack (blue dots), the vanilla scheme is compromised by the random bit flipping  attack (black dots), and the robust scheme  is compromised by the correlation attack (red dots). Each dot represents a different experiment. Dots above the attack success boundary (green line) represent a successful attack (in which the database owner incorrectly blames an honest SP for the unauthorized sharing).}\label{fig:performance_all_scheme}
  %\vspace{-7.5mm}
\end{figure}
% \end{wrapfigure}

We first show the vulnerability of the vanilla scheme under the   correlation attacks. 
To change sufficient number of database entries during fingerprint insertion, we let both row- and column-wise fingerprint density (i.e., $\gamma_r$ and $\gamma_l$) vary in \{0.05, 0.06, 0.07, 0.08, 0.09, 0.1\}, which gives 36 different fingerprinted databases. Then, we let these databases be compromised by $\mathrm{Atk_{row}(\mathcal{S})}$ followed by  $\mathrm{Atk_{col}}(\mathcal{J})$. For each compromised database, we record its percentage of changed entries (i.e., $\mathrm{Per_{chg}} = 1-Acc$) caused by the correlation attack as well as the resulting fingerprint robustness measured in terms of $\mathrm{Per_{cmp}}$ (percentage of compromised fingerprint bits), and scatter the results as blue dots in Figure \ref{fig:performance_all_scheme} (we will discuss the black and red dots in the figure in later experiments). As discussed in Section \ref{sec:robustness_metrics} and empirically shown in \cite{ji_fp_raid2021}, as long as the malicious SP can compromise more than $50\%$ fingerprint bits, it is able to avoid being detected as the traitor and cause the database owner to accuse other innocent SPs who have also received the database. Thus, we say an attack is successful if $\mathrm{Per_{cmp}}>50\%$, and the green horizontal line in Figure \ref{fig:performance_all_scheme} represents the attack success  boundary. 
From Figure \ref{fig:performance_all_scheme}, we observe that  in most of the cases, the identified correlation attacks can compromise more than $50\%$ fingerprint bits (blue points that are above the green line) at the cost of changing only less $5\%$ SNPs in the vanilla fingerprinted database (i.e., an attacker can distort the majority of the fingerprint bits by also keeping the utility of the database high).

\begin{table}
% \begin{wraptable}{r}{.5\linewidth}
%\vspace{-4mm}
    \begin{minipage}{0.9\linewidth}
\begin{adjustbox}{width=1\textwidth}
\begin{tabular}{|l|c|c|c|c|c|c|c|}\hline
\diagbox{p-value\\ consistency}{$\gamma_r=\gamma_l$}&
  0.05 & 0.06 & 0.07 & 0.08 & 0.09 & 0.1\\ 
  \hline
  vanilla scheme& 100\%&  98\%&    98\%&    98\%&    98\%&    96\%  \\ 
  \hline
  vanilla after corr. attacks  & 98\%&    96\%&    96\%&    96\%&    96\%&    92\%  \\
  \hline
    \hline
  robust scheme & 92\%&    94\%&    94\%&    88\%&    94\%&    92\%\\
  \hline
  robust after corr. attacks & 84\%&    92\%&    84\%&    80\%&    86\%&    82\%  \\
%   \hline
%   0.09 & 33.59\% &    33.59\% &    34.38\% &    34.38\% &    34.38\% &  35.16\%\\
%   \hline
%   0.1 & 27.34\% &    27.34\% &    27.34\% &    27.34\% &    28.12\% &    28.91\%  \\
    \hline
  \end{tabular}%\vspace{-6mm}
  \end{adjustbox}
   \end{minipage}%\vspace{-4mm}
    \caption{Consistency of SNP-phenotype association compared with the ground-truth.}\label{table:p-value}
%\vspace{-6mm}
% \end{wraptable}
\end{table}

In Table \ref{table:p-value}, we show the consistency of SNP-phenotype association study (discussed in Section \ref{sec:utility_metrics}) after the vanilla fingerprinting and the correlation attacks. In particular, we first obtain the set of top-50 SNPs having strong associations with a phenotype (i.e., the 50 SNPs with the lowest p-values) from the original (non-fingerprinted) database and denote this set as the ground-truth set.  Next, we  get the new sets of top-50 SNPs from the vanilla fingerprinted database and the correlation attack-compromised database. Then, we evaluate the consistency by counting the percentage of overlapping SNPs between them and the ground-truth set. From  the upper panel of  Table \ref{table:p-value}, we observe that the vanilla fingerprinting preserves high utility of the consistency and the correlation attacks on vanilla fingerprinting scheme also maintain such utility.  For example, one of the successful attacks happens when $\gamma_r = \gamma_l=0.09$ (blue dot indicated by the blue arrow in Figure \ref{fig:performance_all_scheme}), and 
% after conducting the correlation attacks on the vanilla fingerprinted database, 
the resultant pirated database copy still preserves more than 96\% SNP-phenotype association.% when $\gamma_r = \gamma_l=0.09$.

% \begin{table}[htb]
% \begin{center}
% \begin{tabular}{|l|c|c|c|c|c|c|c|}\hline 
% \diagbox{p-value\\ consistency}{$\gamma_r=\gamma_l$}&
%   0.05 & 0.06 & 0.07 & 0.08 & 0.09 & 0.1\\ 
%   \hline
%   vanilla scheme& 100\%&  98\%&    98\%&    98\%&    98\%&    96\%  \\ 
%   \hline
%   vanilla after   attacks  & 98\%&    96\%&    96\%&    96\%&    96\%&    92\%  \\
%   \hline
%       \hline
%   robust scheme & 92\%&    94\%&    94\%&    88\%&    94\%&    92\%\\
%   \hline
%   robust after   attacks & 84\%&    92\%&    84\%&    80\%&    86\%&    82\%  \\
%     \hline
%   \end{tabular}
%       \caption{Consistency of SNP-phenotype association compared with the ground-truth obtained from the original (non-fingerprinted database).}\label{table:p-value}
%       \end{center}
%       \end{table}

    %\vspace{-4mm}
\subsubsection{Performance against Random Bit Flipping Attack}
Next, we compare the attack ability of the random bit flipping attack with our identified correlation attacks. To show the effectiveness of the correlation attacks, 
%make a fair comparison, 
we let the random bit flipping attack change %approximately the same   (or slightly more)
more percentage of entries in the vanilla fingerprinted genomic database than the correlation attacks. 
% In Table \ref{table:vanilla_vs_rnd}, 
We also record the fingerprint robustness after the vanilla fingerprinted database is subject to random bit flipping attack. In particular, we  set $\gamma_r = \gamma_l \in\{0.05, 0.06, 0.07, 0.08, 0.09, 0.1\}$, and let the malicious SP randomly change a certain percentage (i.e., $\mathrm{Per_{chg}}\in\{5\%, 7\%, 9\%,  11\%, 13\% 15\%\}$) of entries in its received copy of the vanilla fingerprinted genomic database.  Thus, we also obtain 36 instance of vanilla fingerprinted databases compromised by random bit flipping attacks. We scatter the recorded results as black dots  in Figure \ref{fig:performance_all_scheme}.  Clearly, the random bit flipping attack can hardly compromise more than half of the fingerprint bits if the database owner inserts dense fingerprint in the genomic database, i.e., when $\gamma_r=\gamma_l\geq 0.06$. In particular, when $\gamma_r=\gamma_l=0.09$, the malicious SP can only distort $35.16\%$ fingerprint bits at the cost of changing   $15\%$ of the SNPs  if it launches the random bit flipping attack (indicated by the black arrow in Figure \ref{fig:performance_all_scheme}). 
In contrast, it can compromise $52.34\%$ of the fingerprint bits at the cost of only changing the values of $3.11\%$ of the SNPs if it launches the correlation attacks  (indicated by the blue arrow in Figure \ref{fig:performance_all_scheme}). This clearly suggests that our vanilla fingerprint scheme developed specially for genomic databases is robust against random bit flipping attacks, but is vulnerable to the attacks using correlations among genome data. Compared with our previous work \cite{ji_fp_raid2021}, the identified correlation attacks for genomic data are even more powerful than the ones identified for a general relational database. The reason is that the correlation patterns in genomic data (e.g., Mendel's law and  linkage disequilibrium) are much stronger than the patterns in a general relational database (i.e., census database in \cite{ji_fp_raid2021}). Thus, the robust fingerprinting for genomic databases is critical.

\subsection{Robust Genomic Database Fingerprinting Against Correlation Attacks}\label{sec:robust_exp}
% %\vspace{-2mm}
% mtg_chg_entry_count =

%     0.0287    0.0289    0.0291    0.0310    0.0302    0.0305
%     0.0289    0.0282    0.0285    0.0298    0.0297    0.0299
%     0.0286    0.0285    0.0292    0.0306    0.0300    0.0303
%     0.0293    0.0295    0.0298    0.0317    0.0309    0.0320
%     0.0287    0.0287    0.0291    0.0313    0.0289    0.0310
%     0.0298    0.0295    0.0307    0.0329    0.0317    0.0336

% \begin{wraptable}{r}{.4\linewidth}

%     \begin{minipage}{1\linewidth}
% \begin{adjustbox}{width=1\textwidth}

Now, we investigate the impact of the proposed  robust genomic database fingerprinting scheme. Recall that the robust fingerprinting  is achieved by post-process the vanilla fingerprinted genomic database using two mitigation techniques, i.e., $\mathrm{Mtg_{row}(\mathcal{S})}$ and   $\mathrm{Mtg_{col}}(\mathcal{J})$.

%\vspace{-2mm}
\subsubsection{Impact on Database Utility}
% \begin{wraptable}{r}{.45\linewidth}
\begin{table}
%\vspace{-8mm}
    \begin{minipage}{0.9\linewidth}
\begin{adjustbox}{width=1\textwidth}
\begin{tabular}{|l|c|c|c|c|c|c|}\hline
\diagbox{$\gamma_r$}{$\gamma_l$}&
  0.05 & 0.06 & 0.07 & 0.08 & 0.09 &0.1\\ \hline
  0.05& 2.87\% &    2.89\% &    2.91\% &    3.10\% &    3.02\% &    3.05\% \\ \hline
  0.06 & 2.89\% &    2.82\% &    2.85\% &    2.98\% &         2.97\% &         2.99\%\\
  \hline
  0.07 & 2.86\% &    2.85\% &    2.92\% &    3.06\% &    3.00\% &    3.03\%   \\
  \hline
  0.08 & 2.93\% &    2.95\% &    2.98\% &    3.17\% &    3.09\% &    3.20\%  \\
  \hline
  0.09 &2.87\% &    2.87\% &    2.91\% &    3.13\% &    2.89\% &    3.10\%  \\
  \hline
  0.1 & 2.98\% &    2.95\% &    3.07\% &    3.29\% &    3.17\% &    3.36\%   \\
  \hline
  \end{tabular}%\vspace{-6mm}
  \end{adjustbox}
   \end{minipage}
   %\vspace{-6mm}
\caption{Additional change caused by   mitigation.}\label{table:chg_mtg}
%\vspace{-7mm}
% \end{wraptable}
\end{table}

In Table \ref{table:chg_mtg}, we record the additional percentage of entries being changed due to the post-processing steps ($\mathrm{Mtg_{row}(\mathcal{S})}$ and   $\mathrm{Mtg_{col}}(\mathcal{J})$). % (recall that the robust scheme is developed based on the vanilla scheme by incorporating two mitigation techniques as the post-processing steps).
Clearly, as shown in Table \ref{table:chg_mtg}, the mitigation  techniques only need to change about $3\%$ of the SNPs in order to make the post-processed database has row-wise and column-wise correlation close to $\mathcal{S}$ and $\mathcal{J}$ and at the same time comply with the Mendel's law. Thus, the robust fingerprint scheme can preserve high utility of the genomic database, i.e., the database accuracy and     the consistency of SNP-phenotype association. For example, as shown in the lower panel of  Table \ref{table:p-value},   in most of the cases, robust scheme achieves more than $90\%$ top-50 SNPs match with the original database.

% We first observe that the proposed robust genomic database fingerprinting scheme  can effectively thwart the identified correlation attacks by just changing about $3\%$ of the SNPs in the post-processing steps 

% Besides, the developed mitigation techniques also preserve the consistency of SNP-phenotype association. As shown in Table \ref{table:p-value} (the row corresponding to the robust scheme), in most of the cases, robust scheme achieves more than $90\%$ top-50 SNPs matches with the original database. 

% \begin{table}[htb]
% \centering
% \begin{tabular}{|l|c|c|c|c|c|c|}\hline
% \diagbox{$\gamma_r$}{$\gamma_l$}&
%   0.05 & 0.06 & 0.07 & 0.08 & 0.09 &0.1\\ \hline
%   0.05& 2.87\% &    2.89\% &    2.91\% &    3.10\% &    3.02\% &    3.05\% \\ \hline
%   0.06 & 2.89\% &    2.82\% &    2.85\% &    2.98\% &         2.97\% &         2.99\%\\
%   \hline
%   0.07 & 2.86\% &    2.85\% &    2.92\% &    3.06\% &    3.00\% &    3.03\%   \\
%   \hline
%   0.08 & 2.93\% &    2.95\% &    2.98\% &    3.17\% &    3.09\% &    3.20\%  \\
%   \hline
%   0.09 &2.87\% &    2.87\% &    2.91\% &    3.13\% &    2.89\% &    3.10\%  \\
%   \hline
%   0.1 & 2.98\% &    2.95\% &    3.07\% &    3.29\% &    3.17\% &    3.36\%   \\
%   \hline
%   \end{tabular}
% %   \end{adjustbox}
% %   \end{minipage}
%     \caption{Additional change caused by the mitigation techniques.}\label{table:chg_mtg}
%     % %\vspace{-6mm}
% % \end{wraptable} 
% \end{table}

\subsubsection{Impact on Fingerprinting Robustness}

% In Table \ref{table:robust_robust_corr} and \ref{table:error_robust_corr}, 

In Figure \ref{fig:performance_all_scheme}, using red dots, we scatter the  fingerprint robustness and the percentage of changed entries when the robust scheme is under the identified correlation attacks (which are identical with that considered in Section \ref{sec:vanilla_exp}). In particular, comparing with the blue dots, we see that the number of successful attacks (red dots above the green line) is significantly reduced by the proposed mitigation techniques. This suggests that  it is very difficult for the malicious SP to compromise more than half of the fingerprint bits using the correlation attacks under the proposed robust scheme, even if the malicious SP has changed more than 20\% of the SNPs in the received database. 
% If the malicious SP wants to achieve attack performance that is close to that on the vanilla scheme (e.g., compromising close to 50\% fingerprint bits), it needs to change a significant amount of SNPs (e.g., about 30\%). 
Moreover, correlation attacks on top of the robust fingerprinted genomic database also significantly reduce the utility of SNP-phenotype association study. As shown in Table \ref{table:p-value} (the row corresponding to robust after correlation  attacks), the percentage of matched top-50 SNPs drops more than $10\%$ compared to the original database.  This   suggests that   the proposed robust genomic database fingerprinting scheme  can effectively thwart the identified correlation attacks by just changing about $3\%$ of the SNPs in the post-processing steps and at the same time maintain high database accuracy and consistency of SNP-phenotype association.

\subsection{Scalability}
\label{sec:large}
\begin{figure}[!htb]
%\vspace{-4mm}
% \minipage{0.23\textwidth}
\centering
  \includegraphics[width=0.77\linewidth]{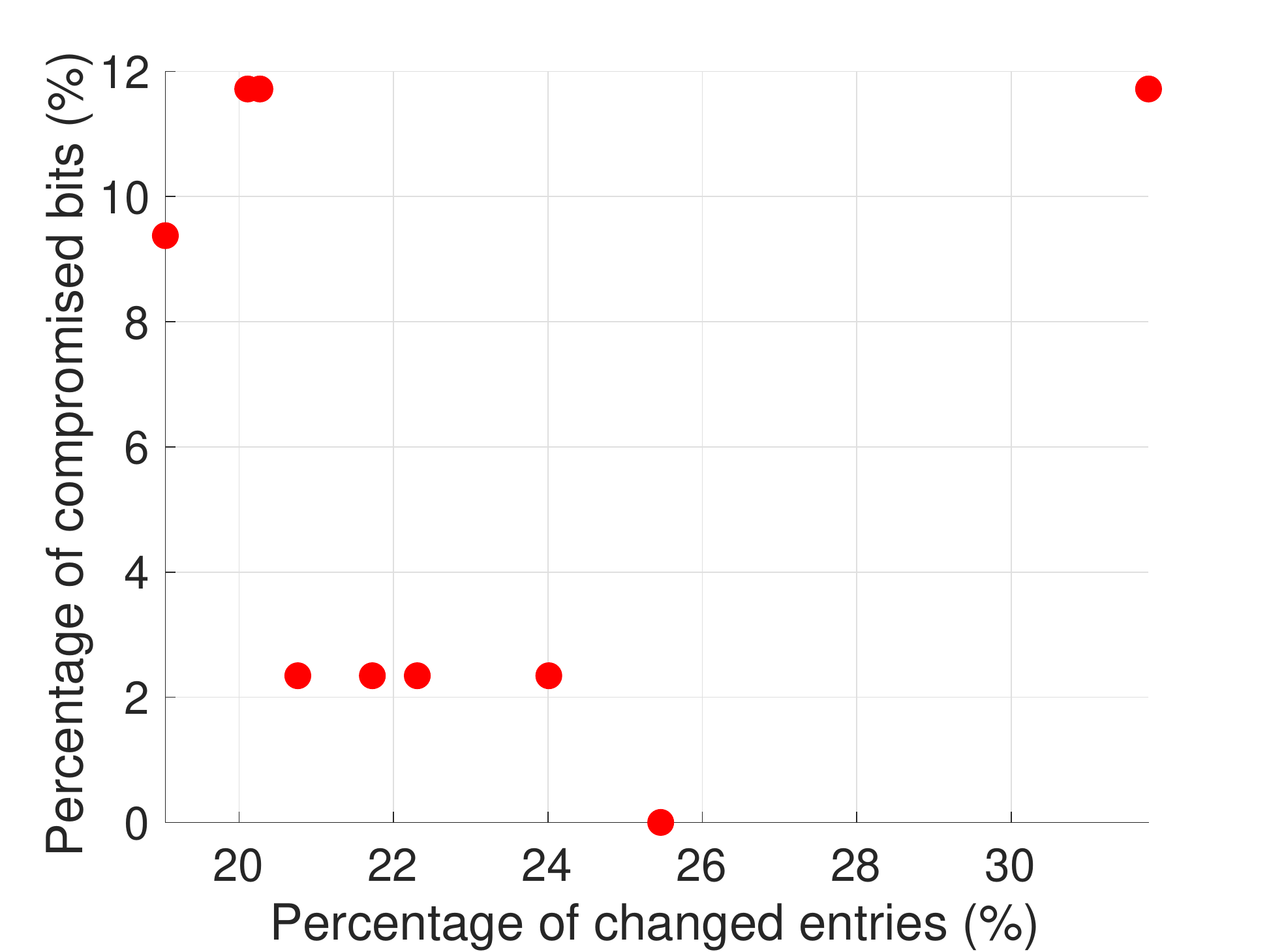}
  %\vspace{-6mm}
  \caption{Fingerprint robustness versus utility loss when the proposed robust scheme is compromised by the correlation attacks.}\label{fig:per_cmp_large}
  \end{figure}

\begin{figure}[!htb]
\centering
  \includegraphics[width=0.77\linewidth]{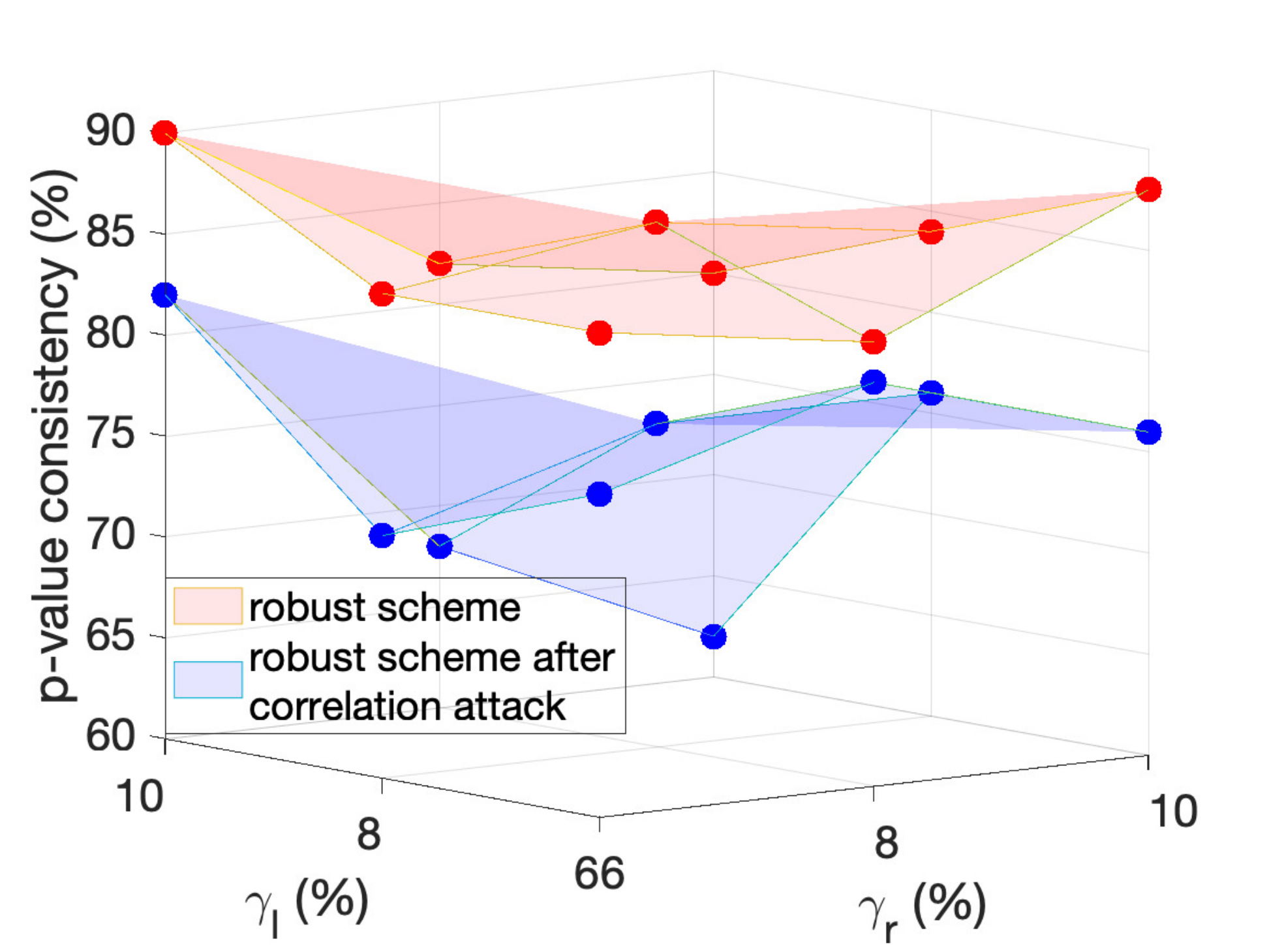}
  %\vspace{-6mm}
    \caption{p-value consistency before and after the  proposed robust scheme is compromised by the correlation attacks.}\label{fig:p_cons_large}
% \endminipage
% %\vspace{-4mm}
\end{figure}

% %\vspace{-2mm}
Now, we investigate the performance of the proposed robust  fingerprinting scheme for larger genomic databases, where each individual has a higher number of SNPs (i.e., 234). In particular, we consider 8000 individuals  among which there are 1333 families (due to the same reasoning in Section \ref{sec:dataset}, we also focus on the correlation between mother-child-father tuple).   %Since the original dataset~\cite{gibbs2003international} only contains 156 SNPs for each individual, we increase the length to 234 by replicating half of the SNP sequence for each person.  
In this experiment, we let $\gamma_r=\gamma_l\in\{0.06, 0.08, 0.1\}$. 
We scatter the pair of percentage
of changed entries and percentage of compromised fingerprint bits in Figure \ref{fig:per_cmp_large}.
We also plot the p-value consistency before and after the robust scheme is subject to the identified correlation attacks in Figure \ref{fig:p_cons_large}. 
From Figure \ref{fig:per_cmp_large}, we see that the robustness increases as the database increases (in terms of both rows and columns), because the correlation attacks cannot distort more than 12\% of the fingerprint bits even though more than 20\% entries are modified. Figure \ref{fig:p_cons_large} further suggests that if the malicious SP launches the correlation attacks on a robust fingerprinted genomic database, the p-value consistency will drop by 10\% on average. This experiment shows that our proposed robust  fingerprinting scheme is also promising when sharing large genomic databases.

% \endminipage
%\vspace{-4mm}
% \hfill
%\vspace{-7mm}
% \minipage{0.23\textwidth}

\section{Discussion}\label{sec:discussion}

\noindent\textbf{Independent treatment of elements in SNP sequence.} Note that by checking the change of  inner product before and after fingerprinting, (\ref{eq:row_defense}) essentially treats each element in the SNP sequence independently. However, in practice, blocks of SNP elements may also contain inherent structure, e.g., two or more individuals are identical by descent (IBD)    if they have inherited blocks of SNPs from a common ancestor without genetic recombination. Thus, a malicious SP may also use this  structural information during an  attack. In future work, we will extend the proposed robust fingerprinting scheme to incorporate the recombination of  IBD segments during meiosis.

\noindent\textbf{Limited side-effect of row-wise mitigation.} In the row-wise mitigation, we post-process each pair of first degree family members   in a family set. This may    impact the similarity of other pairs in which either individual is involved. 
For example, updating mother-child pair may increase the similarity of grandfather-grandchild pair. However, as discussed in Section \ref{sec:dataset}, such impact is very limited for higher degree family members.

\noindent\textbf{Assumption on  prior knowledge.} To the advantage of the malicious SP, we assume that it  has  at least equally accurate knowledge about the genomic database (i.e., Mendel's law, row-wise and column-wise correlation)  compared with the database owner. We do not consider specific  auxiliary information (such as SNP population frequencies, rare disease-associated variants, population stratification, and SNP-phenotype associations) in this paper. If the malicious SP has more auxiliary information than the database owner (which rarely happens in   real world applications), the robustness of the proposed scheme may be compromised.  Such robustness degradation will be limited for generic relational database, i.e., the malicious SP still cannot distort more than half of the fingerprint bits  \cite{ji_fp_raid2021}. We will empirically investigate this for genomic databases  by considering various case studies   in the future work.

\noindent\textbf{Privacy concerns during genomic database sharing.} The primary goal of database fingerprinting is to claim copyright and prevent unauthorized redistribution, however, privacy concerns and regulations may also impede genomic data sharing. In our   recent work \cite{ji2021differentially}, we   developed a novel scheme which can leverage the  intrinsic randomness introduced by fingerprinting to provide provable privacy guarantees in relational database sharing, i.e., copyright and privacy protection can be achieved in one shot.   In future, we will also study privacy-preserving genomic database fingerprinting by adapting the scheme in \cite{ji2021differentially}.

\noindent\textbf{Computational complexity 0f mitigation techniques.} If the genomic database contains $M$ individuals, each of which has $N$ SNPs, then the computation complexity for $\mathrm{Mtg_{row}}(\mathcal{S})$ is $\mathcal{O}(MN)$, because solving (\ref{eq:row_defense}) requires checking all SNPs of each mother-child-father tuple. The  computation complexity for   $\mathrm{Mtg_{col}}(\mathcal{J})$ is $\mathcal{O}(\frac{3N}{\alpha})$, where 3 is the number of possible instances of 
SNP  values and $\alpha$ is the desired error in Sinkhorn-based optimal transport \cite{le2021robust}.

%\vspace{-7mm}

\section{Conclusion}
\label{sec:conclusion}
%\vspace{-1mm}

In this paper, we have proposed robust fingerprinting for genomic  databases composed of SNP sequences. To this end, we first  identified the row-wise and column-wise correlation attack which utilize   Mendel’s  law  and  linkage  disequilibrium to distort the embedded fingerprint bits. Next, we   developed a vanilla fingerprinting scheme specifically for genomic database by allowing the database owner to embed more fingerprint in each selected SNP sequence. Then, we further made this vanilla scheme robust against the identified correlation attacks by augmenting it with two mitigation techniques, which serve as post-processing steps for the vanilla scheme. In particular, the row-wise mitigation is achieved via solving a  cumulative absolute distance minimization, and the  column-wise mitigation is realized using optimal mass transport  of distributions. Via experiments, we have shown that the identified correlation attacks are much more powerful than common attacks against fingerprinting schemes; they can easily distort more than half of the fingerprint bits at a small cost of database utility. However, these attacks are effectively alleviated by our developed mitigation techniques. The proposed scheme has the potential to further motivate researchers to share their genomic databases with each other, knowing that the shared database is of high utility and the recipient will be hesitant to leak the database due to the provided liability guarantees via the proposed robust fingerprinting scheme.
%\vspace{-4mm}

% First, we have validated  the vulnerability of existing database fingerprinting schemes by identifying different correlation attacks: column-wise correlation attack (which utilizes the joint distributions among attributes), row-wise correlation attack (which utilizes the statistical relationships among the rows), and integration of them. Next, to defend against the identified attacks, we have  developed mitigation techniques that can work as post-processing steps for any off-the-shelf database fingerprinting schemes. Specifically, the column-wise mitigation technique modifies limited entries in the fingerprinted database by solving a set of optimal mass transportation problems concerning pairs of marginal distributions. On the other hand, the row-wise  mitigation technique modifies a small fraction of the fingerprinted database entries by solving a combinatorial search problem. We have also empirically investigated the impact of the identified correlation attacks and the performance of proposed mitigation techniques on two real-world relational databases. Experimental results show high success rates for the correlation attacks and high robustness for the proposed mitigation techniques, which alleviate the correlation attacks even using inaccurate/noisy data correlations.  

%\section*{Acknowledgement}
%Research reported in this publication was supported by the National Library Of Medicine of the National Institutes of Health under Award Number R01LM013429.

\typeout{}
\bibliographystyle{plain}
\bibliography{fingerprinting.bib}

\end{document}